\documentclass{aastex631}
\usepackage{graphicx}  
\usepackage{dcolumn}   
\usepackage{bm}        
\usepackage{amssymb}   

\usepackage{xcolor}
\usepackage{amsmath}

\newcommand{\be}{\begin{equation}}
\newcommand{\ee}{\end{equation}}
\newcommand{\beq}{\begin{eqnarray}}
\newcommand{\eeq}{\end{eqnarray}}

\def\numu{\mathrel{{\nu_\mu}}}

\def\barnumu{\mathrel{{\bar \nu}_\mu}}

\def\t13{\mathrel{{\theta_{13}}}}
\def\y12{\mathrel{{\tan^2 \theta_{12}}}}
\def\c2{\mathrel{{\chi^2 }}}
\def\msun{\mathrel{{M_\odot }}}

\def \sun	 	{$_{\odot}$}

\newcommand{\eg}{{\it e.g.}}

\newcommand{\eq}{Eq.}

\newcommand{\Fig}{Fig.}

\newcommand{\Sec}{Sec.}

\newcommand{\App}{App.}

\newcommand{\Tab}{Table}

\newcommand{\equ}[1]{\eq~(\ref{equ:#1})}
\newcommand{\figu}[1]{\Fig~\ref{fig:#1}}

\newcommand{\n}{neutrino}
\newcommand{\ns}{neutrinos}
\newcommand{\ic}{IceCube}
\newcommand{\td}{TDE}
\newcommand{\bh}{SMBH}
\newcommand{\tds}{TDEs}

\newcommand{\dsg}{AT2019\texttt{dsg}}
\newcommand{\fdr}{AT2019\texttt{fdr}}
\newcommand{\aalc}{AT2019\texttt{aalc}}

\shorttitle{Neutrino emission from three TDEs}
\shortauthors{Winter, Lunardini}

\begin{document}

\title{Interpretation of the observed neutrino emission from three Tidal Disruption Events}

\author[0000-0001-7062-0289]{Walter Winter}
\affiliation{Deutsches Elektronen-Synchrotron DESY, Platanenallee 6, 15738 Zeuthen, Germany} 

\author[0000-0002-9253-1663]{Cecilia Lunardini}
\affiliation{Department of Physics, Arizona State University,  450 E. Tyler Mall, Tempe, AZ 85287-1504 USA}

\date{\today}

\begin{abstract}
Three Tidal Disruption Event (TDE) candidates (\dsg , \fdr , \aalc) have been associated with high energy astrophysical neutrinos in multi-messenger follow-ups. In all cases, the \n\ observation occurred  ${\mathcal O}(100)$ days after the maximum of the optical-ultraviolet (OUV) luminosity.  
We discuss unified fully time-dependent interpretations of the neutrino signals where the neutrino delays are not a statistical effect, but rather the consequence of a physical scale of the post-disruption system.
Noting that X-rays flares and infrared (IR) dust echoes  have been observed in all cases, we 
consider three models in which quasi-isotropic neutrino emission is due to the interactions of accelerated protons of moderate, medium, and ultra-high 
energies with  X-rays, OUV, and IR photons, respectively.   We find that  the neutrino time delays can be well described in the X-ray model assuming magnetic confinement of protons in a calorimetric approach if the unobscured X-ray luminosity is roughly constant over time, and in the IR model, where the delay is directly correlated with the time evolution of the echo luminosity (for which a model is developed here). 
The OUV model exhibits the highest neutrino production efficiency. In all three models, the highest neutrino fluence is predicted for  \aalc, due to its high estimated supermassive black hole mass and low redshift. All models result in diffuse \n\ fluxes that are consistent with observations.
\end{abstract}



\section{Introduction}


Nearly a decade after their discovery, the  high-energy extragalactic neutrinos seen by IceCube~\citep{IceCube:2013low} -- possibly indicating the production sites of the Ultra-High Energy Cosmic Rays (UHECRs) -- are still largely a mystery, as their origin is still unresolved. Neutrino alert-triggered follow-up searches in electromagnetic data have proven successful to identify individual Active Galactic Nuclei (AGN) blazars as sources; the most prominent case is TXS 0506+056, which was found to be in a gamma-ray flaring state during the neutrino emission~\citep{IceCube:2018dnn}. In time-integrated point source searches, individual neutrino sources are also emerging~\citep{IceCube:2019cia}: three AGN blazars (including TXS 0506+056), and a starburst galaxy (NGC 1068). The most sensitive limits for transient sources exist for the stacking of Gamma-Ray Bursts~\citep{IceCube:2012qza,IceCube:2017amx}, which indicate that these can only contribute to the diffuse neutrino flux at the percent level. Arguments from actual neutrino event detections and population statistics~\citep{Bartos:2021tok}, as well as from spectral shape and directional information~\citep{Palladino:2018evm} point towards multiple source populations contributing to the astrophysical diffuse neutrino flux;  such populations could be (apart from mis-identified atmospheric neutrino events) AGN blazars, AGN cores, starburst galaxies, neutrinos of Galactic origin, and Tidal Disruption Events (TDEs). 


\tds\ are phenomena in which a massive star passes close enough to a supermassive black hole (SMBH) to be ripped apart by its tidal forces. Following this process of tidal disruption, 
about half of the star's matter remains bound to the \bh\ and is ultimately accreted onto it.  Observationally, this mass accretion results in a months- or year-long flare, with the emission of photons over a wide range of wavelengths, including  a  black body (BB) spectrum in the optical-ultraviolet (OUV) range, as well as sometimes X-ray, infrared (IR) and radio emission, see \eg\ \citet{Stein:2020xhk}. From observations and numerical modeling, the basic picture of the post-accretion phase of a \td\ has emerged, including an accretion disk, a semi-relativistic outflow, and possibly a jet, see \eg\ \citet{Dai:2018jbr}. Neutrinos  have been associated with TDEs through follow-up searches; the Zwicky Transient Facility (ZTF) has been especially successful, leading to the identification of \dsg\ \citep{Stein:2020xhk}  and \fdr\ \citep{Reusch:2021ztx} as optical counterparts of two neutrinos (IceCube events IC191001A and IC200530A respectively). Afterwards, it was noticed that  these \tds\ were accompanied by an echo due to reprocessing of BB and X-ray radiation into the IR by surrounding dust, and this \n-dust link then led to the identification of a 
third \td, \aalc, as counterpart of the IceCube event IC191119A~\citep{vanVelzen:2021zsm}. 
With three neutrino-TDE associations in less than one year, the case for \tds\ as \n\ sources has become stronger\footnote{Note that \fdr\ and \aalc\ have not uniquely been identified as TDEs. Alternative interpretations are, \eg\ AGN accretion flares or even luminous supernovae~\citep{Pitik:2021dyf}. All events share  TDE-characteristic features, namely the evolution of the BB light curve -- including the large and rapid optical flux increase~\citep{Reusch:2021ztx} -- and the large dust echoes~\citep{vanVelzen:2021zsm}. Also note that \fdr\ and \aalc\ occurred in AGN (i.e., black holes that were accreting prior to the optical flare). However, the extreme properties of the flares compared to normal AGN variability suggest that the optical outbursts are likely induced by the disruption of a star. Here we adopt the hypothesis that the three objects considered here are indeed \tds, and therefore will be called as such; the wording ``candidates'' will be dropped from here on.}, and it is therefore timely to revisit the neutrino production mechanism in \tds, and the  contribution of \tds\ to the observed \n\ diffuse flux at \ic, which was constrained to be $\lesssim$30\% in a stacking search~\citep{Stein:2019ivm}.


Neutrino production in TDEs was proposed earlier in jetted models~\citep{Wang:2011ip,Wang:2015mmh,Dai:2016gtz,Lunardini:2016xwi,Senno:2016bso}, mostly motivated by observations of the jetted TDE Swift J1644+57. Furthermore, neutrino production in different disk states \citep{Hayasaki:2019kjy} and ejecta-external medium interactions \citep{Fang:2020bkm} were considered. TDEs may also be candidates to accelerate and even power the UHECRs~\citep{Farrar:2008ex,Farrar:2014yla,Guepin:2017abw,Biehl:2017hnb,Zhang:2017hom}. For \dsg , jets \citep{Winter:2020ptf,Liu:2020isi}, outflow-cloud interactions~\citep{Wu:2021vaw}, disk, corona, hidden winds or jets \citep{Murase:2020lnu} have been proposed, see  \citet{Hayasaki:2021jem} for an overview. While a collimated outflow, such as a jet, has the advantage that it can provide the necessary power for the neutrino emission (see discussion in~\citet{Winter:2021lyo}), no convincing direct jet signatures for \dsg\ have been observed~\citep{Mohan:2021flu}, and the observed radio signal might only be interpreted as jet signature in scenarios with purely leptonic radiative signatures for an unnaturally narrow jet~\citep{Cendes:2021bvp} or a steep density profile~\citep{Cannizzaro:2020xzc}.
For \fdr,  corona, hidden wind and jet models have been considered in \citet{Reusch:2021ztx}, and in \citet{vanVelzen:2021zsm} a disk model for all three \tds\ has been proposed. The neutrino production site is therefore uncertain, and comparative quantitative studies of all three \tds\ do not yet exist.\footnote{After completion of this work, a choked jet model for all three \tds\ has been presented in \citet{Zheng:2022kam}.}


In this work, we provide a unified quantitative description of the three observed \n-emitting \tds, \dsg , \fdr , and \aalc. We build on the fact that these \tds\ have a few common characteristics  beyond the detected IR dust echoes: (i) the most likely neutrino energies are in the 100~TeV range, and (ii) the neutrinos arrived ${\mathcal O}(100)$ days after the BB peak -- when the BB luminosities have already decreased significantly, but the dust echoes have been close to their maxima. Moreover, (iii) X-rays from all the neutrino-associated sources have been detected, although X-ray detection is generally rare in \tds, see \eg\ \citet{vanVelzen:2020cwu}. 
In all cases, 
(iv) the estimated SMBH masses  (with large uncertainties) are  between about $10^{6.5}$ and $10^{7.5} \, M_\odot$~\citep{vanVelzen:2021zsm}, with two of them being higher than the mean of the observed population ($\overline M \simeq 10^{6.57}~\msun$, see \cite{Nicholl:2022pmn,Ramsden:2022oxt}). 
Consequently, all events should have correspondingly high BB luminosities as a consequence of high Eddington luminosities -- for which the measured values in the OUV range are only lower limits, due to obscuration effects.

These commonalities immediately raise important questions: is the neutrino production associated with the X-ray, OUV, or IR signals, i.e., what is the smoking gun signature for the neutrino production? What causes the neutrino time delay with respect to the BB peak, and is that always expected? What can we learn from the predicted neutrino spectra in comparison with the observed neutrino energies? Are the neutrino spectra evolving with time?
Some of these questions have been examined in a qualitative way, for example by suggesting that the neutrino time delay might be directly related to the time evolution of the mass accretion rate, which could be delayed by the debris circularization time, or stay constant over a time scale of hundreds of days \citep{vanVelzen:2021zsm}.  Here we address these questions by developing a fully quantitative, time-dependent model of neutrino production. We take the point of view that the neutrino time delays have a physical origin -- in a characteristic time or length scale of the post-disruption system -- instead of being statistical effects.
We 
aim at keeping the models as minimal as possible, by only introducing the strictly necessary ingredients that can be common to different acceleration scenarios. Specific possible accelerators and their feasibility will be discussed briefly for context. 

Our study is organized as follows: in \Sec~\ref{sec:model} we introduce the model and describe its details. Results are presented for three realizations (named M-X, M-OUV and M-IR, after the three different photon targets used), in \Sec~\ref{sec:mx}, \Sec~\ref{sec:mouv}, and \Sec~\ref{sec:mir}.
We present our results for the diffuses fluxes in \Sec~\ref{sec:diffuse}, and we compare the different TDEs and models in the comparison and discussion section \Sec~\ref{sec:discussion}. We finally summarize in \Sec~\ref{sec:summary}. Three technical appendices are included at the end of the paper.

\section{Model description}
\label{sec:model}

\begin{figure}[t]
    \centering
    \includegraphics[width=0.6\textwidth]{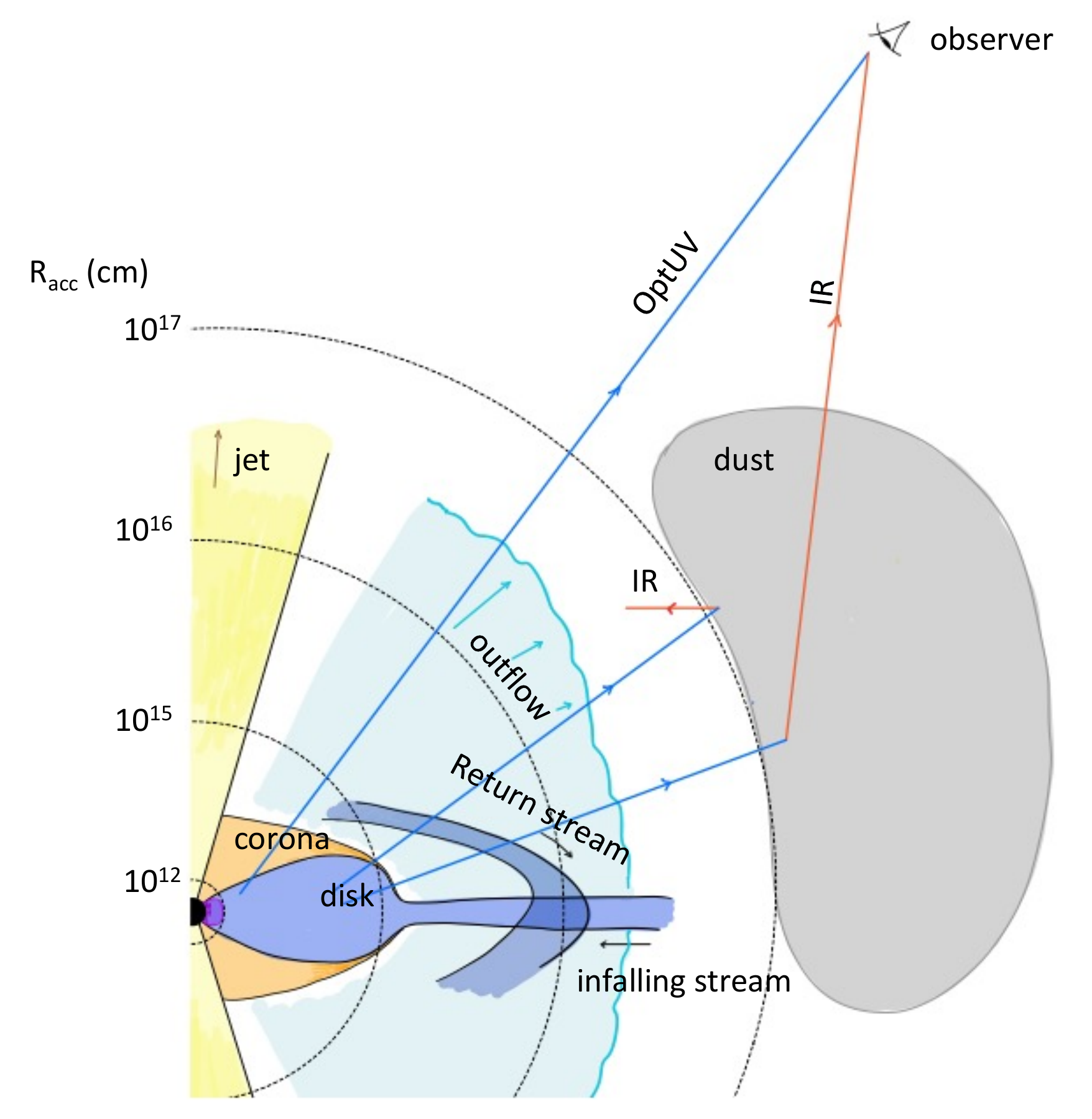}
    \caption{Global geometry of a TDE, illustrated. We show the main elements, some of which may not always be present (\eg\ the jet). Three photon histories are sketched, representing the OUV flare, the delayed IR echo, and backwards-emitted IR photons. The dotted circles indicate the several possibilities for the acceleration radius, $R_{\mathrm{acc}}$ (the smallest scale, corresponding to the X-ray photosphere inside the accretion disk, is expanded for visibility). A reference  \bh\ mass $M=10^7 \, \msun$  is adopted here. }
    \label{fig:macro_cartoon}
\end{figure}

In this section, we describe the spirit and the ingredients of our model. For a reader wanting a quick overview, \Sec~\ref{sec:overview}, \figu{macro_cartoon} and \Tab~\ref{tab:observations} summarize the qualitative features and numbers. Our method to describe the dust echo is presented in \Sec~\ref{sec:dust}.

\subsection{Overview}
\label{sec:overview}

To explain the emission of high energy \ns, protons must be accelerated to energies beyond $\sim$PeV, and interact with background photons and/or matter. The \ns\ are then natural results of these interactions.
In the spirit of minimality, we assume that the non-thermal proton injection: (i) is (quasi-)isotropic (rather than beamed as in jetted models), and (ii) evolves in time like the mass accretion rate. The latter assumption seem contrary to the idea of reproducing the \n\ time delays, because it naturally leads to a neutrino signal that follows the BB luminosity evolution. Still, within this basic scenario, there are multiple ways to reproduce the \n\ delays, since these can be related to other important time or length scales of the post-disruption system. 

Elaborating on the latter point, let us examine the large scale structure of a \td\ after the disruption has occurred, as shown in \figu{macro_cartoon}. In the figure, the components where proton acceleration could take place, and their characteristic radial scales, are illustrated for a reference  \bh\ mass $M=10^7 \, \msun$ (gravitational radius $R_S \simeq 3 \cdot 10^{12} \, \mathrm{cm}$). The figure illustrates how the location of the proton acceleration (radial distance $R=R_{\mathrm{acc}}$)  can vary widely, from  $R_{\mathrm{acc}}\sim (3-30) \, R_S$ (the X-ray photosphere and the hot corona, see, e.g., \citet{Murase:2020lnu} and discussion in  \citet{vanVelzen:2021zsm}), to  $R_{\mathrm{acc}}\sim 10^3 \, R_S$ (the OUV photosphere, or the collision region inside a jet, see, e.g., \citet{Dai:2018jbr}) or even larger values like  $R_{\mathrm{acc}}\sim 10^{16} -10^{17} \, \mathrm{cm}$, for acceleration inside the outflow (possibly near the dust torus) \citep{Stein:2020xhk}, or in 
stream-stream collisions \citep{Dai:2015eua,Hayasaki:2019kjy}.
Which of these possibilities are consistent with our models will be examined  later (see Secs. \ref{sec:mx}, \ref{sec:mouv} and \ref{sec:mir}). 
In some cases, the \n\ delay might be reproduced by a geometric distance; for example higher-energy protons could interact with the IR photons from the dust echo, which carries a delay from the size of the dust region. In other cases, the delay might arise from the dynamics of proton propagation:  \eg\ lower-energy protons -- such as those leaking out of an off-axis jet (see \App~\ref{app:jetted} for a discussion) -- could be confined in magnetic fields over the diffusion time, and transfer energy into secondaries in the calorimetric limit. For generality, here 
we do not specify the {\em proton acceleration zone} in detail, but rather characterize it by the maximal proton energy achieved,  $E_{p,\mathrm{max}}$ (equivalent to the maximal Lorentz factor), and the proton injection luminosity and its evolution, similarly to AGN blazar radiation models, see e.g. \citet{Keivani:2018rnh,Gao:2018mnu,Oikonomou:2022gtz}. For a discussion on realistic proton energies that can be achieved if the acceleration and radiative processes are considered, as well as other constraints (for shock-acceleration, the shocks ought not be radiation-mediated), see  \App~\ref{app:maxproton}. We focus instead on the \emph{radiation zone}, i.e., the region at $R \gtrsim R_{\mathrm{acc}}$  where the neutrinos are produced. 

We start by observing 
that the preferred photon target depends on the maximal proton energy provided by the accelerator.
Indeed, for $p\gamma$ interactions, the observed neutrino energies in the 100 TeV range indicate black body target photon temperatures\footnote{In this estimate, both the pitch-angle averaged cross section (see Fig. 4 in \citet{Hummer:2010vx}) and the peak of the photon number density located at $\simeq 2.8 \, T$~\citep{Fiorillo:2021hty} are taken into account.}
\begin{equation}
T  \simeq  80 \, \mathrm{eV} \left(\frac{E_\nu }{100 \,\mathrm{TeV}} \right)^{-1}~,
\label{equ:target}
\end{equation}
in the (soft) X-ray range. Since, however, depending on the spectral shape the actual neutrino energies may be significantly higher for observed muon tracks, lower target photon temperatures in the optical-UV (OUV) or infrared (IR) ranges paired with neutrinos peaking at higher energies may work as well, provided that the accelerator is efficient enough. Using \equ{target}, the requirement
\begin{equation}
E_{p,\mathrm{max}} \gtrsim 20 \, E_\nu \simeq 160 \, \mathrm{PeV}\left(\frac{T }{\mathrm{eV}} \right)^{-1}~
\label{equ:epmax}
\end{equation}
translates then into $E_{p,\mathrm{max}} \gtrsim 2 \, \mathrm{PeV}$, $E_{p,\mathrm{max}} \gtrsim 100 \, \mathrm{PeV}$, and $E_{p,\mathrm{max}} \gtrsim 1 \, \mathrm{EeV}$ for X-rays, OUV, and IR targets, respectively. Detailed parameters for the individual TDE electromagnetic spectra are listed in \Tab~\ref{tab:observations}. We will study these options systematically, increasing  $E_{p,\mathrm{max}}$ within three models called M-X, M-OUV, and M-IR, making additional target photon fields accessible for the interactions. In some cases, $pp$ interactions with the outflow will also contribute significantly, especially if no high target photon densities are accessible. 

Note that our model assumptions will be as universal as possible, which means that common parameters are chosen for all TDEs if physically motivated. This assumption simplifies the comparison, whereas the very different redshifts of the TDE candidates with associated neutrinos jeopardize the direct comparison of neutrino fluences or event rates, see \Tab~\ref{tab:observations}; in fact, we will see that the predicted neutrino emission is in fact not as different as one may expect if the neutrino luminosity or spectra (at the source) are compared. 

Compared to other models in the literature, our model M-X shares some similarities with the hidden wind model in \citet{Murase:2020lnu}, and our model M-OUV is in fact a time-dependent numerical implementation of the idea proposed in~\citet{Stein:2020xhk}. Our model M-IR is motivated by the dust echo connection of the neutrino-emitting TDEs~\citep{vanVelzen:2021zsm}, postulating a direct connection. 
Therefore, we develop our own dust model to obtain the time-dependent luminosity of the IR echo (see \Sec~\ref{sec:dust}). 
Compared to jetted models, the main challenge for the presented models (and in fact most quasi-isotropic emission models) are very high required transfer efficiencies of material of the disrupted star into non-thermal protons; this problem can be avoided in models with a collimated outflow, see discussion in~\citet{Winter:2021lyo}. However, the difference between isotropic and collimated emission models is more fundamental: are TDEs inefficient neutrino emitters each with the contribution of a fraction of an event on average, invoking the Eddington bias argument~\citep{Strotjohann:2018ufz}, or are they more efficient neutrino emitters perhaps just not pointing into our direction or are they too far away in most cases? Since we focus on the isotropic case in this study, the predicted neutrino event number per TDE will be an interesting indicator if the Eddington bias argument has to be invoked for each individual TDE.

In the following subsections, we describe the elements that are common to all the models, whereas specifics will be discussed in the following respective results sections. 

\begin{table}[t]
    \centering
    \begin{tabular}{|l|ccc|}
    \hline
     & {\bf \dsg} & {\bf \fdr} & {\bf \aalc} \\
     \hline
    \multicolumn{4}{|l|}{{\bf Overall parameters}} \\
    \hline
    Redshift $z$ &  0.051 (1) & 0.267 (2) & 0.036 (3) \\
    $t_{\mathrm{peak}}$ (MJD) & 58603 (4) & 58675 (2)\footnote{Peak position uncertain; here a value close to epoch~1 of the BB peak in (2) is chosen.}  & 58658 (3) \\
    SMBH mass $M$ [$M_\odot$] & $5.0 \, 10^6$ (3) & $1.3 \, 10^7$ (3) & $1.6 \, 10^7$ (3)
     \\
    \hline
    \multicolumn{4}{|l|}{{\bf Neutrino observations} } \\
    \hline
    Name (includes $t_\nu$) & IC191001A (5) & IC200530A (6) & IC191119A (7) \\
    $t_\nu-t_{\mathrm{peak}}$ [days] &  154  &  324  & 148  \\
    $E_\nu$ [TeV] & 217 (5) & 82 (6)  & 176 (7) \\
    $N_\nu$ (expected, GFU) &  0.008--0.76 (1) &  0.007--0.13 (2) & not available  \\
    \hline
    \multicolumn{4}{|l|}{{\bf Black body (OUV)}} \\
    \hline
    $T_{\mathrm{BB}}$ [eV] at $t_{\mathrm{peak}}$ & 3.4 (1) &  1.2 (2) & 0.9 [\Sec~\ref{sec:dust}] \\
        $L_{\mathrm{BB}}^{\mathrm{bol}}$ (min.) [$\frac{\text{erg}}{\text{s}}$] at $t_{\mathrm{peak}}$ & 
        $2.8 \, 10^{44}$ (\Sec~\ref{sec:dust}) &
       $1.4 \, 10^{45}$ (\Sec~\ref{sec:dust}) &
        $2.7 \, 10^{44}$ (\Sec~\ref{sec:dust}) \\
            BB evolution from & (1)  & (2) & (3) \\
        \hline
     \multicolumn{4}{|l|}{{\bf X-rays (X)}} \\
     \hline
        $T_{\mathrm{X}}$ [eV]  &  72 (1) & 56 (2,3) & 172 (3) \\
        $L_{\mathrm{X}}^{\mathrm{bol}}$ [$\frac{\text{erg}}{\text{s}}$] @ $t-t_{\mathrm{peak}}$ &  $6.2 \, 10^{43}$ @ 17$\, \text{d}$ (1) & $6.4 \, 10^{43}$  @ 609$\, \text{d}$  
        (2) & $1.6 \, 10^{42}$  @ 495$\, \text{d}$ 
        (3)  \\
    \hline
     \multicolumn{4}{|l|}{{\bf Dust echo (IR) }} \\
     \hline
 $T_{\mathrm{IR}}$ [eV]  &  0.16 (\Sec~\ref{sec:dust}) & 0.15~(2) &  0.16 (\Sec~\ref{sec:dust})  \\
 Time delay $\Delta t$ [d]  & 239 (\Sec~\ref{sec:dust}) & 155 (\Sec~\ref{sec:dust}) & 78 (\Sec~\ref{sec:dust}) \\
 $L_{\mathrm{IR}}^{\mathrm{bol}}$ [$\frac{\text{erg}}{\text{s}}$] @ $t-t_{\mathrm{peak}}$ 
 & $2.8 \, 10^{43}$ @ 431$\, \text{d}$ (\Sec~\ref{sec:dust})  & $5.2 \, 10^{44}$ @ 277$\, \text{d}$  (\Sec~\ref{sec:dust})  &  $1.1 \, 10^{44}$ @ 123$\, \text{d}$ (\Sec~\ref{sec:dust}) \\
     \hline
     \multicolumn{4}{|l|}{{\bf Universal model assumptions -- and their consequences }} \\
     \hline
  $\varepsilon_{\mathrm{diss}}$ ($L_p/\dot M$) & $0.05-0.2$   & $0.05-0.2$ & $0.05-0.2$  \\
        $F_{\mathrm{peak}}$ ($\dot M/L_{\mathrm{edd}}$ at  $t_{\mathrm{peak}}$) & $100$  & $100$  & $100$ \\
        $M_\star/M_\odot$ ($M_\star \simeq 2 \cdot \int \dot M dt$) & $0.6$  & $5.7$  & $6.3$ \\
     $t_{\mathrm{dyn}}$ [days] (interval with $\dot M \gtrsim L_{\mathrm{edd}}$) & $670$  & $1730$  & $1970$  \\
     \hline
    \end{tabular}
    \caption{Summary of observations and universal model ingredients; references to the original articles or sections in this article are given as well in brackets: (1) \citet{Stein:2020xhk}, (2) \citet{Reusch:2021ztx}, (3) \citet{vanVelzen:2021zsm}, (4) \citet{vanVelzen:2020cwu}, (5) \citet{2019GCN.25913....1I}, (6) \citet{IC200530A}, (7) \citet{2019GCN.26258....1I}. The X-ray and IR luminosities are given at the indicated times, with the evolution determined by our theoretical models (for details, see main text). The neutrino time delay $t_\nu-t_{\mathrm{peak}}$  is computed from $t_{\mathrm{peak}}$ and $t_\nu$. See caption of \figu{mx} for the definition of the GFU (Gamma-Ray Follow-Up)  effective area.  }
    \label{tab:observations}
\end{table}

\subsection{Numerical time-dependent simulation of radiation zone}
 
We solve the coupled differential equation system for the in-source densities $N_i$ (differential in energy and volume)
\begin{equation}
\frac{\partial N_{i}(E,t)}{\partial t}=\underbrace{\tilde J_{i}\left(E ,  t\right)}_{\text{Injection}}+ \underbrace{\frac{\partial}{\partial E}\left(\frac{E}{t_{\mathrm{cool}}(E,t)} \, N_{i}\left(E, t\right)\right)}_{\text{Cooling}}-\underbrace{\frac{N_{i}\left(E,t\right)}{t_{\text{esc}} \left(E,t\right)}}_{\text{Escape}}  \label{equ:kinetic}
\end{equation}
for the protons ($i=p$) and neutrons ($i=n$) in a fully time-dependent way using the NeuCosmA software~\citep{Hummer:2010vx,Hummer:2011ms,Boncioli:2016lkt,Biehl:2017zlw}. Here the cooling rate is given by $t_{\text{cool}}^{-1} = E^{-1} \left| dE/dt \right|$, the escape rate by $t^{-1}_{\text{esc}}$, and the injection rate $\tilde J_p \left(E\right)=J_p(E)+J_{j \rightarrow p}\left( E \right)$ (differential in volume, energy and time) contains the injection from the acceleration zone $J_p$ as well as the re-injection $J_{j \rightarrow p}$ from interacting protons at higher energies and interacting neutrons ($j=p,n$) -- which couples the differential equations; for neutrons corresponding terms are used, but there is no injection from an acceleration zone, i.e., $J_n (E) \equiv 0$. 

Photohadronic interactions are treated as discrete energy losses as described in~\citet{Biehl:2017zlw}, which means that escape terms with the interaction rate $t^{-1}_{p \gamma}$ are added and the interaction products are re-injected in $J_{j \rightarrow p}$ at lower energies; we use the efficient but accurate  treatment of \citet{Biehl:2017zlw,Hummer:2010vx} based in the physics of SOPHIA~\citep{Mucke:1999yb}. The emitted neutrinos are integrated over time, while the in-source densities $N_i$ are evolved over the lifetime of the system (as shown in our figures). Since the proton injection rate varies with time, a steady state will never be reached. Protons also cool via Bethe Heitler pair production; note that all these injection, escape and cooling rates are explicitly time-dependent through the time-dependent evolution of the target photon fields.

Protons escape diffusively, see below, or by interactions, whereas neutrons escape over the free-streaming timescale $t_{\mathrm{fs}} \simeq R/c$. We also assume that protons escape over the duration of the TDE, referred to as dynamical timescale $t_{\mathrm{dyn}}$ here (see \Sec~\ref{sec:energetics} for its definition as one possible measure for the duration of the TDE event) to take into account the transient nature of the event; however, the impact of that term on the neutrino production is small. We also include  synchrotron losses for all charged species (including the secondary muons, pions, kaons).\footnote{For the neutrino flavor composition, the effects are negligible because of very high critical energies due to the relatively small values of $B$; see \eg\ App. A.1 of \citet{Baerwald:2011ee}. }

Our approach can properly treat the optically thick case (see App.~C of \citet{Biehl:2017zlw}), which we define as $\tau_{p\gamma}^{\mathrm{fs}} \equiv t_{\mathrm{fs}}/t_{p \gamma} > 1$ (protons interact efficiently while crossing the radiation zone), and the calorimetric case, which we define as $\tau_{p\gamma}^{\mathrm{cal}} \equiv t_{\mathrm{dyn}}/t_{p \gamma} > 1$ (magnetically confined protons interact efficiently over the duration of the system, even if $\tau_{p\gamma}^{\mathrm{fs}}<1$). 
We will find the calorimetric case for interactions in model M-X, where the proton energies are low, and the optically thick case for M-OUV, where the interaction rates are high (M-IR is in between). Note that the effective photohadronic cooling rate $t^{-1}_{p \gamma, \mathrm{cool}} \simeq 0.2 \, t^{-1}_{p \gamma}$ does not apply to the (optically thin) calorimetric case, since neutrons produced in the interactions can escape over the free-streaming timescale, and hence the effective cooling rate will be much higher; consequently, we only show interaction rates where applicable, while our numerical treatment  reproduces all these effects self-consistently. 
Furthermore, note that the calorimetric approach requires that adiabatic cooling is sufficiently small, which depends on the expansion of the radiation zone: if (magnetically confined) protons lose energy faster than they can interact, the adiabatic cooling  will affect the production rate of neutrinos especially for model M-X; see discussion
in \App~\ref{app:adiabatic}.

\subsection{Energetics, proton injection, and proton confinement}
\label{sec:energetics}

The Eddington luminosity 
$ L_{\mathrm{edd}} \simeq 1.26 \, 10^{45} \, \left( M/(10^7 \, M_\odot )\right) \, \mathrm{erg \, s^{-1}}  $ -- where $M$ is the SMBH mass --  is a measure for the energy re-processing rate through the SMBH, where super-Eddington mass accretion rates are expected for TDEs at peak; we use the inferred SMBH masses listed in \Tab~\ref{tab:observations}.  Following \citet{Dai:2018jbr}, we assume that the mass accretion rate $\dot M$ exceeds $L_{\mathrm{edd}}$ by a factor $F_{\mathrm{peak}} \simeq 100$ at peak. While the mass fallback rate is expected to scale $\propto t^{-5/3}$ generically, we more accurately implement that the mass accretion rate roughly follows the observed BB evolution for each individual TDE; for details and references, see \Tab~\ref{tab:observations}. 

We parameterize the proton injection spectrum $J_p$ (density differential in energy and time) as
\begin{equation}
 J_p = J_0 \, E_p^{-2} \, \exp \left( - \frac{E_p}{E_{p,\mathrm{max}}} \right) \, ,     \label{equ:inj}
\end{equation}
where $E_{p,\mathrm{max}}$ is a parameter depending on the model (M-X, M-OUV, or M-IR). The non-thermal proton injection luminosity
$L_p$ is dynamically following the mass accretion rate $\dot M$, and is given by
\begin{equation}
L_p = \varepsilon_{\mathrm{diss}} \, \dot M \, c^2 =  \varepsilon_{\mathrm{Comp}} \, \varepsilon_{\mathrm{NT}} \, \dot M \, c^2 \, .
\label{equ:Lp}
\end{equation}
Here $\varepsilon_{\mathrm{diss}} \equiv  \varepsilon_{\mathrm{Comp}} \, \varepsilon_{\mathrm{NT}}$ is the dissipation efficiency, which we define as the fraction of the accretion luminosity, $\dot M \, c^2$, which is converted into non-thermal particles (which are dominated by protons in our models).
It contains the conversion efficiency, $\varepsilon_{\mathrm{Comp}}$, from the infalling material into a component, such as an outflow, corona, or jet; for example, in the TDE unified model~\citet{Dai:2018jbr}, $\varepsilon_{\mathrm{Comp}} \simeq 0.2$ for outflow or jet. It also contains the dissipation efficiency of kinetic power into non-thermal particles, $\varepsilon_{\mathrm{NT}}$.  
For the outflow and wind models, one may estimate that $\varepsilon_{\mathrm{NT}} \sim v^2/c^2$, whereas for jetted internal shock models (see e.g.  Eq. (10) in \citet{Daigne:1998xc} or Eq. (4) in \citet{Rudolph:2019ccl} for the definition of $\varepsilon_{\mathrm {NT}}$ in this case) one finds a wide range $\varepsilon_{\mathrm{NT}} \sim 0.1-0.4$, see \App~\ref{app:jetted}. 
We need to postulate relatively high overall dissipation efficiencies, $\varepsilon_{\mathrm{diss}} \simeq 0.05-0.2$, into non-thermal protons, where $\varepsilon_{\mathrm{diss}} \simeq 0.2$ represents an optimistic choice with  $\varepsilon_{\mathrm{NT}} \sim 1$, and  $\varepsilon_{\mathrm{diss}} \simeq 0.05$ corresponds to an outflow with $v=0.5 \, c$ (in the direction of the poles, where the particle acceleration will be most efficient, see \citet{Dai:2018jbr}) or a jet with  $\varepsilon_{\mathrm{NT}} \simeq 0.25$.
We therefore show the results for the range $\varepsilon_{\mathrm{diss}} \simeq 0.05-0.2$, where applicable. Note that, as we will demonstrate, a high value of $\varepsilon_{\mathrm{diss}}$ can be also viewed as a requirement to produce a neutrino fluence compatible with the expected average neutrino event rate derived from observations (see \Tab~\ref{tab:observations}), as the predicted neutrino event rate is proportional to $\varepsilon_{\mathrm{diss}}$.  On the other hand, there are factors which may relax this requirement: we use 1~GeV as lower energy of the non-thermal proton spectrum, as we anticipate that the protons are picked up from a thermal bath; this assumption for the minimal proton energy is conservative, as a lot of energy will be transferred into non-thermal protons which are below the $p\gamma$ threshold; it may result in a reduction of $\varepsilon_{\mathrm{diss}}$ by up to a factor of five if the minimal proton injection energy is higher. Furthermore, note that TDEs that occurred in a AGN -- such as \fdr\ and \aalc\ -- may draw material from the existing disk as, in general, TDEs in AGN seem to be more luminous; see e.g. \citet{Chan:2021blg}. This would also lower the requirement for the dissipation efficiency for a fixed mass of the disrupted star. 

The proton spectrum normalization $J_0$ in the radiation zone  is obtained from the proton luminosity $L_p$ 
\begin{equation}
   \int\limits_{1 \, \mathrm{GeV}}^{\infty} dE_p \, E_p \, J_p(E_p) = \frac{L_p}{\frac{4}{3} \, R^3 \, \pi} \, .
\end{equation} 
We note that the relation between  the size of the radiation zone $R$ and the acceleration location $R_{\mathrm{acc}} \lesssim R$ depends on the model, see next sections. 

Another measure for the available energy  is the mass of the disrupted star. Since about half of the stellar debris accreted towards the SMBH, we can estimate the mass of the disrupted star as
as $M_\star \simeq 2 \cdot \int \dot M dt $,
which we list in terms of solar masses ($M_\odot  \simeq 1.8 \, 10^{54} \, \mathrm{erg}$) as a result of our computation in \Tab~\ref{tab:observations}.\footnote{Integration ranges in time will correspond to the ranges shown in our figures, \eg\ \figu{mx}, upper row. If material from the pre-existing AGN is accreted, half of the reported $M_\star$ correspond to the accreted mass. } 

Let us now introduce other fundamental ingredients of our models. One of them is the  dynamical timescale $t_{\mathrm{dyn}}$ of the TDE, which is roughly estimated as the time period for which $\dot M > L_{\mathrm{edd}}$.  This a suitable definition, since qualitative changes are expected to occur when $\dot M$ becomes sub-Eddington, $\dot M \ll L_{\mathrm{edd}}$ (\eg\ transition of the disk accretion state or jet cessation). In practice, the values of $t_{\mathrm{dyn}}$ are determined numerically by imposing the condition $\dot M > L_{\mathrm{edd}}$ on curves that are derived from the BB light curves, assuming that $\dot M$ evolves like the BB luminosities.\footnote{
 To visualize this exercise, see, e.g., \figu{mx}, where the curves (thick green in upper panels) for $L_p \sim 0.2 \, \dot M$ (for $\varepsilon_{\mathrm{diss}}=0.2$) and $L_{\mathrm{edd}}$ are shown: $t_{\mathrm{dyn}}$ corresponds to the time window during that $\dot M \simeq 5 \, L_p > L_{\mathrm{edd}}$. As it can be also seen from the figures, we resorted to extrapolating the actual light curves due to lack of data covering the entire time period. We stress that precise values of $t_{\mathrm{dyn}}$ do not influence our results much, but are useful as one possible measure for the duration of the TDE. They are of the order 1000 days, with some dependence on the individual TDE.}

If the magnetic field in the radiation zone is given by  $B$, protons gyrate with  the Larmor radius 
\begin{equation}
R_L \simeq 3.3 \, 10^{12} \, \mathrm{cm} \,\left( \frac{E_p}{\mathrm{PeV}} \right) \left(\frac{B}{\mathrm{G}} \right)^{-1} \, .
\label{equ:larmor}
\end{equation}
Note that the confinement condition $R_{\mathrm{acc}} > R_L$ can impose constraints on the size of the accelerator for high enough proton energies; for example,  $R_{\mathrm{acc}} \gtrsim 3 \, 10^{16} \, \mathrm{cm}$ for $E_{p,\mathrm{max}} \simeq 1 \, \mathrm{EeV}$ and $B=0.1 \, \mathrm{G}$ (see model M-IR). 
Assuming Bohm-like diffusion with a diffusion coefficient $D \simeq R_L \, c$, 
protons can be displaced by
 \begin{equation}
 R \simeq \sqrt{D \, t_{p,\mathrm{diff}}} = 3 \, 10^{15} \, \mathrm{cm} \, \left(\frac{E_p}{\mathrm{PeV}} \right)^{1/2}
\left(\frac{B}{\mathrm{G}} \right)^{-1/2} 
\left( \frac{t_{\mathrm{dyn}}}{1000 \, \mathrm{days}} \right)^{1/2} \, ,
\label{equ:dx}
\end{equation}
where the diffusion time is set to the dynamical time $t_{p,\mathrm{diff}} \simeq t_{\mathrm{dyn}}$ and \equ{larmor} has been used for $D$.
This means that protons with PeV energies will be magnetically confined by Gauss-scale magnetic fields in a region of size $R \simeq 3 \, 10^{15} \, \mathrm{cm}$ over the lifetime of the TDE, and the system will be calorimetric if $t_{p \gamma} < t_{\mathrm{dyn}}$, see \eg\ model M-X. Since the Larmor radius in \equ{larmor} 
is for our parameters considerably smaller than the region $R$, only protons with very high energies can escape directly (ballistically), and protons with $E_p \sim 1 - 10 \, \mathrm{PeV}$ (relevant for the X-ray interactions) are confined. Here we follow \citet{Baerwald:2013pu} and define a self-consistent escape rate for the protons given by
\begin{equation}
 t_{p,\mathrm{esc}}^{-1} = \min( t_{\mathrm{fs}}^{-1}  , t_{p,\mathrm{diff}}^{-1}  )   \qquad \text{with} \qquad
 t_{p,\mathrm{diff}}^{-1} \equiv \frac{D}{(c \,  t_{\mathrm{fs}})^2} \, , \label{equ:pescape}
\end{equation}
which implies that the diffusive escape rate is limited by the free-streaming rate.\footnote{This definition mimicks the transition between the diffusive and free-streaming escape regimes for magnetic field turbulence coherence lengths $l_c \sim R$, see discussion in~\citet{Becker:new}.}
 Note that magnetic confinement in turbulent magnetic fields also leads to isotropization of the proton-photon pitch-angles (the angles between incoming protons and photons in the interaction frame), if not already isotropized. 

Finally we discuss the choice of Gauss-scale magnetic fields over such a large region $R$. \citet{Dai:2018jbr} obtain a magnetic flux $\simeq 10^{31} \, \mathrm{G} \, \mathrm{cm^2}$ at about $80 \, R_S$, which translates to about 1 Gauss in a distance of $10^{15} \, \mathrm{cm}$. \citet{Stein:2020xhk} (see also \citet{Stein:thesis}) obtain a field of about $B \simeq 0.07 \, \mathrm{G}$ from the radio equipartition analysis at $R \simeq 7 \, 10^{16} \, \mathrm{cm}$ for a radio epoch that was near-contemporaneous to the neutrino detection. Assuming that the field $B \propto 1/R$ for a toroidal configuration, Gauss-scale fields at $10^{15} \, \mathrm{cm}$ are plausible. In the hidden wind model, Gauss-scale magnetic fields are estimated from the kinetic wind luminosity as well~\citep{Murase:2020lnu,Reusch:2021ztx}, in fact up to 30~G are obtained for \fdr . We obtain similarly large values for our models if the plain equipartition argument is applied to the BB luminosity. Note, however,
that the equipartition argument can only be directly applied if the (related) radiation is generated from electromagnetic processes dominated by these B-fields, such as synchrotron emission in the fast-cooling regime. Here the observed thermal spectra are not related to such processes (compared to e.g. radio emission from an outflow). Since it is crucial for our calorimetric (confinement) models that the magnetic field is high enough (i.e. higher magnetic fields help the confinement), we chose the Gauss-scale as conservative value that meets the confinement requirement. Higher values for the magnetic field would lead to qualitatively similar results.

\subsection{Photon and proton targets}

Protons encounter target photons and protons within the radiation zone of radius $R$. All photon targets are described by quasi-thermal spectra with temperature $T$, motivated by observations. As a short summary, 
for X-rays, which stem from the accretion disk, we hypothesize that the (highest) detected X-ray flux is indicative for the actually emitted X-ray flux, and obscuration, such as from a complicated geometry or outflow, leads to flux fluctuations -- whereas the intrinsic flux within $R$ is relatively stable (see \eg\ \citet{Wen:2020cpm}). For OUV, we take the BB evolution from observations directly, but we correct for absorption by inferring the unabsorbed luminosity from the dust echo, see \Sec~\ref{sec:dust}. For IR, we model the dust echo both in terms of time-dependence and normalization, as described in the same subsection. More details will be also given in the respective model sections later. The in-source photon density $n(\varepsilon)$ (typically units [GeV$^{-1}$ cm$^{-3}$]) is then computed from the luminosity $L$ as
\begin{equation}
   \int d \varepsilon \, \varepsilon \, n(\varepsilon) = \frac{L \, \tau}{4 \, \pi \,  R^2 \, c} \, ,
   \label{equ:norm}
\end{equation}
were $\tau$ is the photon optical thickness approximated by $\tau \sim 1$ (and the implied effective escape time is $t_{\mathrm{esc}} \equiv \tau \cdot R/(3 \, c)$).  This is a good estimate for X-rays ($\tau \simeq 1$), a lower limit estimate for the OUV black body if $R < R_{\mathrm{BB}}$ ($\tau > 1$) and a better estimate if $R \gtrsim R_{\mathrm{BB}}$ ($\tau \simeq 1$), and a rough estimate  for the IR target if $R$ corresponds to the scale of the dust scattering.

We  also consider $pp$ interactions with a mildly relativistic outflow because the outflow is a plausible model ingredient, as it may be the reason for the X-ray obscuration. Moreover, the outflow is expected in numerical simulations \eg~\citet{Dai:2018jbr}, and it has been directly observed for \dsg\ \citep{Stein:2020xhk}. 
Note that there may be interactions of other components, such as debris, clumps or clouds as well, which we however do not describe in view of major geometric and density uncertainties.

As shown in~\citet{Dai:2018jbr}, the outflow densities are high up to about 1000 gravitational radii, which is about $10^{15} \,  \mathrm{cm}$ for $M \simeq 10^7 \, M_\odot$ in consistency with \equ{dx}; therefore even calorimetric effects may be expected. The relevant target density is computed by assuming that a fraction $\varepsilon_{\text{Outflow}} \simeq 0.2$ of the mass accretion is re-processed into the outflow so that $L_{\text{outflow}}(t) = \varepsilon_{\mathrm{outflow}} \, \dot M(t)$ (scaling with the mass accretion rate)~\citep{Dai:2018jbr}. Since $M_{\mathrm{outflow}} \simeq L_{\text{outflow}} \, \frac{c}{v} \, t_{\text{fs}}$ in the production volume, smaller velocities imply higher densities.
The free-streaming optical thickness can be estimated as\footnote{For this estimate, we use $t_{pp}^{-1} \simeq \sigma_{pp} \, n_H \, c$, where $n_H$ is the target density and $\sigma_{pp} \simeq 60 \, \mathrm{mbarn}$ is the cross section for $E_p \simeq 1 \, \mathrm{PeV}$~\citep{Kelner:2006tc}. In  our numerical approach, the full energy-dependence of the cross section is taken into account, following~\citet{Kelner:2006tc}.}
\begin{equation}
 \tau_{p p}^{\mathrm{fs}} \equiv \frac{t_{\mathrm{fs}}}{t_{p p}} \simeq 0.01 \,  \left( \frac{M}{10^{7} \, M_\odot} \right) \left( \frac{R}{10^{15} \, \mathrm{cm}} \right)^{-1} \left( \frac{v}{0.5 \, c} \right)^{-1} \, ,
\label{equ:tppfs}
\end{equation}
where $v$ is the velocity of the outflow. A comparison to~\citet{Dai:2018jbr} reveals that $v \simeq 0.5 \, c$  near the funnel, whereas $v \simeq 0.1 \, c$  perpendicular to that -- where the densities are higher.  We conservatively use $v \simeq 0.5 \, c$.
We will see  that $pp$ interactions can be important if the X-ray luminosity is low, or at low energies (below the $p\gamma$ threshold). Especially in the calorimetric case, the system can be optically thick for $pp$ interactions as well:
\begin{equation}
 \tau_{p p}^{\mathrm{cal}} \equiv \frac{t_{\mathrm{dyn}}}{t_{p p}} \simeq 19 \,  \left( \frac{M}{10^{7} \, M_\odot} \right) \left( \frac{R}{10^{15} \, \mathrm{cm}} \right)^{-2} \left( \frac{v}{0.5 \, c} \right)^{-1} \, \left(  \frac{t_\mathrm{dyn}}{600 \, \mathrm{days}} \right) \, .
\label{equ:tppcal}
\end{equation}

\subsection{Dust echo and inferred bolometric luminosities}
\label{sec:dust}

For each \td, an IR lightcurve was measured at several times after the peak by neoWISE in the W1 and W2 frequency bands. 
It has been interpreted as thermal emission from a dust torus which is illuminated and heated by the OUV and X-ray radiation emitted by the TDE accretion disk (see, \eg, \citet{vanVelzen:2021zsm}). 
The main features of this IR dust echo are: (i) the delayed emission with respect to the primary OUV and X-ray emission, due to the dust being at an angle with respect to line of sight (see Fig. \ref{fig:macro_cartoon}); 
and (ii) a thermal IR spectrum with temperature at or below the dust sublimation temperature of $T_{\text{IR}}\simeq 0.16 \, \text{eV}$ ($\simeq 1850 \, \text{K}$) \citep{Reusch:2021ztx,vanVelzen:2021zsm} (see also \citet{vanVelzen:2016jsk} for a general description of dust properties in \tds). 
We use this value unless $T_{\mathrm IR}$ has been measured, see \Tab~\ref{tab:observations}.

Assuming that the contribution of the X-rays to the dust echo is negligible in the present case, the energy emitted in IR in the neoWISE bands can be expressed as: 
\be
E_{\text{IR}} =\epsilon^{\text{IR}}_{\text{bol}} \, \epsilon_\Omega \, \epsilon_{\text{dust}} E^{\text{bol}}_{\text{BB}} \, .
\label{equ:EIR}
\ee
Here $E^{\text{bol}}_{\text{BB}}$ is the total (bolometric) energy in the OUV spectrum, $\epsilon^{\text{IR}}_{\text{bol}}$ is a correction factor describing the ratio between  the neoWISE measured luminosity and the bolometric luminosity,  $\epsilon_\Omega = \Omega/4 \pi$ is the geometric covering factor of the dust, and $\epsilon_{\text{dust}} \leq 1$ is an efficiency expressing the fraction of the incident radiation that is re-emitted by the dust in the IR. 

As shown in \citet{Reusch:2021ztx},   the time evolution of the IR luminosity for \fdr\ is well described by convolving the observed OUV luminosity with a (normalized) box function $B(t)$ centered at  time $\Delta t$ and having width $2 \, \sigma_t$ (i.e., $B(t)=1/(2 \, \sigma_t)$ if $\Delta t - \sigma_t\leq t \leq \Delta t+\sigma_t$ and $B(t)=0$ elsewhere, with $\sigma_t \le \Delta t$). Such a function models the fact that a wide spread in time delays is expected due to the extended shape of the dust torus, see  \figu{macro_cartoon}; the quantity $\sigma_t$ describes such a spread, with $\Delta t$ being the central value.
Following  \citet{Reusch:2021ztx}, we choose $\sigma_t = \Delta t$, which accounts for a portion of the IR flux to have zero delay due to some dust being along the line of sight. 
The time-differentiated version of \equ{EIR} reads 
\be
L_{\mathrm{IR}}(t) =\epsilon^{\text{IR}}_{\text{bol}} \, \epsilon_\Omega \, \epsilon_{\text{dust}} \int^{+\infty}_{-\infty} L^{\text{bol}}_{\text{BB}}(t^\prime) \,  B(t-t^\prime) \, dt^\prime \, .
\label{equ:LIR}
\ee

We performed a least-squares fit of the IR luminosity measurements at different times (taken from \citet{Reusch:2021ztx,vanVelzen:2021zsm}), with the goal of obtaining: (i) a best-fit IR lightcurve; (ii) an estimate of the unabsorbed OUV luminosity and temperature and (iii) information on the size and geometry of the dust torus.    Results are given in Table \ref{tab:observations} ; below a more detailed description of the methodology is given. 
 
The IR lightcurve was modeled as in \equ{LIR}, with the assumption that the time profile of $L^{\text{bol}}_{\text{BB}}$ is the same as that of the \emph{observed} OUV luminosity, $L_{\text{BB}}$.  
The results of the fit are the time delay, $\Delta t$, and the normalization $E_{\mathrm{IR}}$.  For all three \tds\ the obtained best-fit IR lightcurve is in good agreement with the data. For \fdr, it is consistent with the one shown in \citet{Reusch:2021ztx}. For \dsg, the lightcurve underestimates the earliest-measured IR flux, but fits the later data points very well.\footnote{The discrepancy at early times might be an indication that the simple model in \equ{LIR} is inaccurate. A possible improvement on it could be to use a bi-modal distribution instead of a box function, to account for the presence of dust on \emph{both} sides of the line of sight, with one side being closer to the observer (resulting in a smaller time delay) than the other.} 

Setting the unknown coefficients in \equ{EIR} to optimistic (large) values, where we took $\epsilon_\Omega \epsilon_{\text{dust}}=0.5$, and estimating the correction factor $\epsilon^{\text{IR}}_{\text{bol}}$  for the W1 and W2 bands\footnote{Here we use thermal spectra with either the measured IR temperature (for \fdr) or a theoretically motivated value close to the dust sublimation temperature  (for \dsg\ and \aalc); we find $(\epsilon^{\text{IR}}_{\text{bol}})^{-1}\simeq 1.26,~1.38$ in the two cases (factors correcting the combined W1+W2 luminosity $\nu L_\nu$).}, a minimum value $E_{\mathrm{BB}}^{\mathrm{bol}}$ (min.), and therefore $L_{\mathrm{BB}}^{\mathrm{bol}}$ (min.) $=E_{\mathrm{BB}}^{\mathrm{bol}} \,  \text{(min.)} \cdot L_{\mathrm{BB}}/E_{\mathrm{BB}}$ was obtained.  From that estimate $L_{\mathrm{BB}}^{\mathrm{bol}}$ (min.), the temperature of the OUV spectrum can be inferred from the Stefan-Boltzmann law; the result was adopted as best estimate available for \aalc\ (in the absence of an estimate from the observed spectrum), and served as a consistency check for the remaining two \tds. See \Tab~\ref{tab:observations} for our results, the $L_{\mathrm{BB}}^{\mathrm{bol}}$ (min.) and $L_{\mathrm{IR}}^{\mathrm{bol}}=(\epsilon^{\text{IR}}_{\text{bol}})^{-1} \, L_{\mathrm{IR}}$  light curves are includes in our results figures (\figu{mx}, \figu{ouv}, and \figu{ir}, upper rows) as black dotted and dashed-dotted curves, respectively.

\section{Moderate-energy  protons interacting with X-rays (model M-X)}
\label{sec:mx}

\begin{figure}[t]
    \centering
    \begin{tabular}{cc}
    \includegraphics[width=0.5\textwidth]{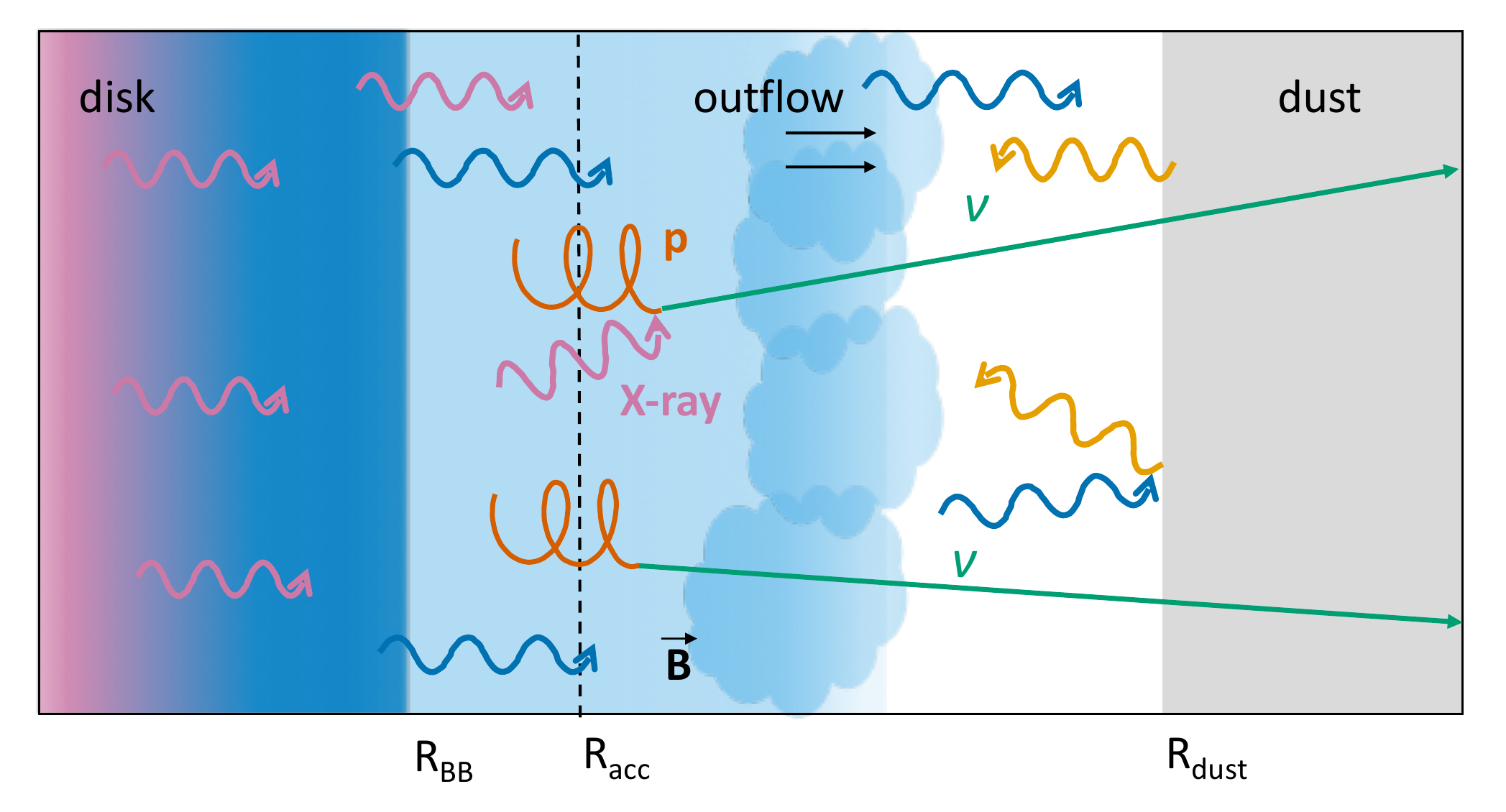}  &  \hspace*{-2cm}
     \raisebox{2.5cm}{
     \begin{tabular}{|l|ccc|}
    \hline
     & {\bf \dsg} & {\bf \fdr} & {\bf \aalc} \\
       \hline
     \multicolumn{4}{|l|}{{\bf Model assumptions}} \\
     \hline
        $R$ [cm] & $5.0 \, 10^{15}$ &  $5.0 \, 10^{15}$ &  $5.0 \, 10^{15}$ \\
        $E_{p,\mathrm{max}}$ [GeV]  & $5.0 \, 10^6$  &
        $5.0 \, 10^6$  &   
        $5.0 \, 10^6$ \\
        $B$ [G] & 1.0 &  1.0 & 1.0   \\
        \hline
     \multicolumn{4}{|l|}{{\bf Model results}} \\
     \hline
     $N_\nu$ GFU & $0.048$  & $0.0048$   &  $0.070$  \\
      & ($0.012$)  & ($0.0012$) &  ($0.018$) \\
     $N_\nu$ PS  & $0.50$  & $0.077$   & $3.3$   \\
      & ($0.12$)  &  ($0.019$) & ($1.8$)  \\
     \hline
    \end{tabular}
    } 
   \end{tabular}
    \caption{Model M-X. Left panel: Microphysics cartoon. Right panel: Model assumptions and results. See caption of \figu{mx} for the definition of the GFU (Gamma-Ray Follow-Up) and PS (Point Source) effective areas. Predicted event rates are for $\varepsilon_{\mathrm{diss}}=0.2$ (0.05). }
    \label{fig:cartoon_MX}
\end{figure}

{\bf Model-specific description.}
Our model M-X uses a low maximal proton energy $E_{p,\mathrm{max}} = 5 \, 10^6 \, \mathrm{GeV}$, universal for all TDEs, which is large enough to guarantee interactions with the X-ray targets in all TDEs, but low enough to suppress the interactions with the OUV; it therefore has the lowest requirement on the proton acceleration efficiency. The microphysics is illustrated in the cartoon in \figu{cartoon_MX}, left panel, and model-specific assumptions and results are listed in the table in \figu{cartoon_MX}, right panel. The radiation zone  is determined by the location of the accelerator, $R \simeq R_{\mathrm{acc}} \gtrsim R_{\mathrm{BB}}$. For the sake of simplicity, we choose $R = 5 \, 10^{15} \, \mathrm{cm}$ together with $B=1 \, \mathrm{G}$ for all three TDEs to satisfy the magnetic confinement condition in \equ{dx}  for the maximal proton energies used here. This means that injected protons will gyrate in magnetic fields and interact with X-rays and protons of the outflow to produce neutrinos, as illustrated in the cartoon.

All three TDEs have in common that they have been observed in X-rays, which are the prime target for the neutrino production of 100~TeV neutrinos, see \equ{target}. However, the X-ray observations have very different characteristics: an exponential early decay (\dsg)~\citep{vanVelzen:2020cwu,Stein:2020xhk}, a late-time observation with strongly varying limits at different times (\fdr)~\citep{Reusch:2021ztx}, and a late-term constant flux (``plateau'') significantly after the neutrino observation (\aalc)~\citep{vanVelzen:2021zsm}. Since X-rays are expected to originate from the accretion disk, the TDE unified model predicts obscuration effects depending on the viewing angle~\citep{Dai:2018jbr}, and large temperature fluctuations on short timescales may be unlikely, we hypothesize that the (highest) detected X-ray signal is indicative for the actually emitted X-ray flux, and that obscuration beyond $R$, such as from a complicated geometry or outflow, leads to the observed fluctuations.\footnote{For \dsg\, the same effect has led to X-ray isotropization of external target photons in the jetted model~\citep{Winter:2020ptf}, where the assumed $R_{\mathrm{acc}} \simeq R_{\mathrm{BB}}$ was somewhat smaller.}  For example, following the slim disk model~\citep{Wen:2020cpm}, the unobscured flux will be relatively stable over time, except when it changes or ceases when the mass accretion rate drops below the Eddington luminosity -- such as if there is a transition of the accretion disk state. Therefore we suppress it  exponentially with a factor $\propto \exp(- L_{\mathrm{edd}}/\dot M)$, which implies that the X-ray flux available for interactions is stable over $t_{\mathrm{dyn}}$. Consequently, we use the measured X-ray temperature $T_X$ for each TDE, and we normalize the spectrum at the time of the highest measured energy flux to the X-ray measurement in the respective energy range -- however we quote bolometric X-ray luminosities in \Tab~\ref{tab:observations}. The in-source photon density $n_X$ can be computed from $L_X$ using \equ{norm}. Note that since $L_X$ hardly changes over time, the neutrino time delay cannot be generated by the X-ray target in our model; 
including fluctuations of $L_X$ would result in corresponding time variations of the neutrino light curve.

\begin{figure}[t]
\begin{center}
\includegraphics[width=0.45\textwidth]{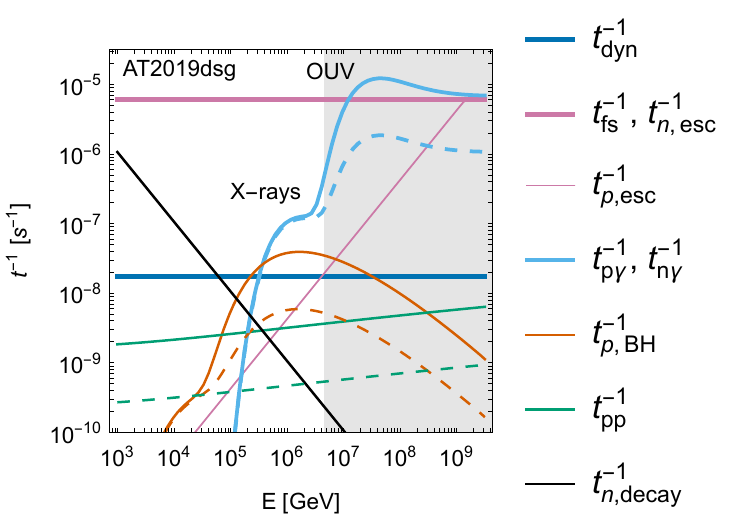} 
\end{center}
\caption{\label{fig:timescales_MX} Relevant inverse timescales (rates) for protons and neutrons for M-X and \dsg\ in the SMBH frame as a function of energy (in the observer's frame); the shaded ares is beyond $E_{p,\mathrm{max}}$ for M-X. Solid curves refer to the $t_{\mathrm{peak}}$, whereas dashed curves to the time of the neutrino emission $t_\nu$ (unless the rate is constant in time). Note that for $p\gamma$ and $pp$ interactions the interaction rates (and not the cooling rates) are shown. }
\end{figure}

{\bf Interaction rates and calorimetric behavior.}
It is useful to look at the rates (inverse timescales)  as a function of energy for one example  to illustrate the calorimetric behavior; therefore we show in \figu{timescales_MX} the rates for \dsg\ at peak time (solid curves) and at the time of the neutrino emission (dashed curves, where time-dependent). First of all, we note that $p \gamma$ interactions with X-rays are possible for $E < E_{p,\mathrm{max}}$, whereas the interactions with the OUV blackbody are suppressed because the proton energies are too low (the gray-shaded area marks $E > E_{p,\mathrm{max}}$). However, since the OUV luminosity is much higher than the X-ray luminosity, and Bethe-Heitler pair production has a lower threshold, the rate of pair production off OUV photons $t^{-1}_{p, \mathrm{BH}}$  (see \Sec~\ref{sec:mouv} for the implementation) can be substantial.\footnote{The effect has also been pointed out in \citet{Murase:2020lnu} for the hidden wind model, specifically.} Depending on the ratio between OUV and X-ray luminosities, X-ray $p\gamma$ interactions may be subdominant. Since the OUV luminosity scales with the mass accretion rate, the corresponding Bethe-Heitler rate will be lower at the neutrino emission time $t_\nu$ -- whereas the X-ray part of $t^{-1}_{p\gamma}$ is quite stable (dashed curves). We observe that while this effect suppresses the overall neutrino production and leads to low neutrino event rates (see \figu{cartoon_MX}, right panel), it favors a late-term neutrino production.\footnote{For \dsg\, the X-ray interaction rate always dominates over $t^{-1}_{p, \mathrm{BH}}$; for \fdr\, X-ray interactions at $t_{\mathrm{peak}}$ are suppressed, but are more efficient at $t_\nu$; for \aalc\, X-ray interactions are always suppressed. } Proton-proton interactions are always relevant for the neutrino production below the $p\gamma$ threshold, but inefficient in the shown example compared to the X-ray interactions; a counter-example is \aalc\, for which  the observed X-ray flux is low.

We can also discuss the proton confinement and calorimetric behavior using \figu{timescales_MX}. Here the free-streaming escape rate $t_{\mathrm{fs}}^{-1}$ is much larger than $t^{-1}_{\mathrm{dyn}}$, and X-ray interactions are effective over the dynamical timescale, but not over the free-streaming timescale. Since protons at the highest energies are confined ($t^{-1}_{p,\mathrm{esc}} \le t^{-1}_{\mathrm{dyn}} \ll t^{-1}_{p\gamma}$), they accumulate, and the in-source proton density will increase. As a result, the $p\gamma$ interactions will be stretched over a longer timescale -- leading to a delay of the neutrino production, scaling with $t_{p \gamma}$.
This can be also seen in analytical estimates: the optical thicknesses for the free-streaming and calorimetric cases can, for the X-rays, be analytically roughly estimated as\footnote{For these analytical estimates, we follow the method in \citet{Guetta:2003wi} using $t^{-1}_{p \gamma} \simeq \sigma_{p \gamma} \, n_X \, c$, where $n_X$ is the target number density and  $\sigma_{p \gamma} \simeq 500 \, \mu \mathrm{barn}$. The photon number density is estimated from the photon energy density divided by the peak energy  of $2.8 \, T_X$~\citep{Fiorillo:2021hty} (where the number density peaks); the estimate therefore only applies to the spectral peak. Compared to numerical computations, for which we follow \citet{Hummer:2010vx}, and which take into account the full-energy dependence of the cross section and the pitch-angle averaging, the optical thickness is typically slightly overestimated (by a factor of a few) in the analytical case because of spectral effects and the neglected width of the $\Delta$-resonance/pitch-angle averaging~\citep{Hummer:2011ms}. }
\begin{eqnarray}
\tau_{p \gamma}^{\mathrm{fs}} &&\equiv \frac{t_{\mathrm{fs}}}{t_{p \gamma}} \simeq 0.06 \,\left( \frac{L_X}{10^{44} \, \mathrm{erg \, s^{-1}}} \right) \left( \frac{T_X}{100 \, \mathrm{eV}} \right)^{-1} \left( \frac{R}{5 \, 10^{15} \, \mathrm{cm}} \right)^{-1}~, \nonumber \\
 \tau_{p \gamma}^{\mathrm{cal}} &&\equiv \frac{t_{\mathrm{dyn}}}{t_{p \gamma}} \simeq 18 \,
\left( \frac{L_X}{10^{44} \, \mathrm{erg \, s^{-1}}} \right) \left( \frac{T_X}{100 \, \mathrm{eV}} \right)^{-1} \left( \frac{R}{5 \, 10^{15} \, \mathrm{cm}} \right)^{-2}
\left( \frac{t_{\mathrm{dyn}}}{600 \, \mathrm{days}} \right)~,
\label{equ:taumx}
\end{eqnarray}
which means that $\tau_{p \gamma}^{\mathrm{fs}} < 1$, but $\tau_{p \gamma}^{\mathrm{cal}}>1$ --  so the system is optically thin, but calorimetric. A small subtlety are neutrons produced in $p\gamma$ interactions, for which free-streaming escape dominates over interactions and decays ($t^{-1}_{n, \mathrm{decay}}= m_n/(E \tau_0)$ with an assumed rest frame lifetime $\tau_0  \simeq 885 \, \mathrm{s}$; see black line in \figu{timescales_MX}); the effective proton cooling rate is therefore closer to the (shown) interaction rate than $0.2 \, t^{-1}_{p \gamma}$. Our numerical code treats all these effects self-consistently, as pointed out earlier.

\begin{figure}[t]
\begin{center}
\includegraphics[width=\textwidth]{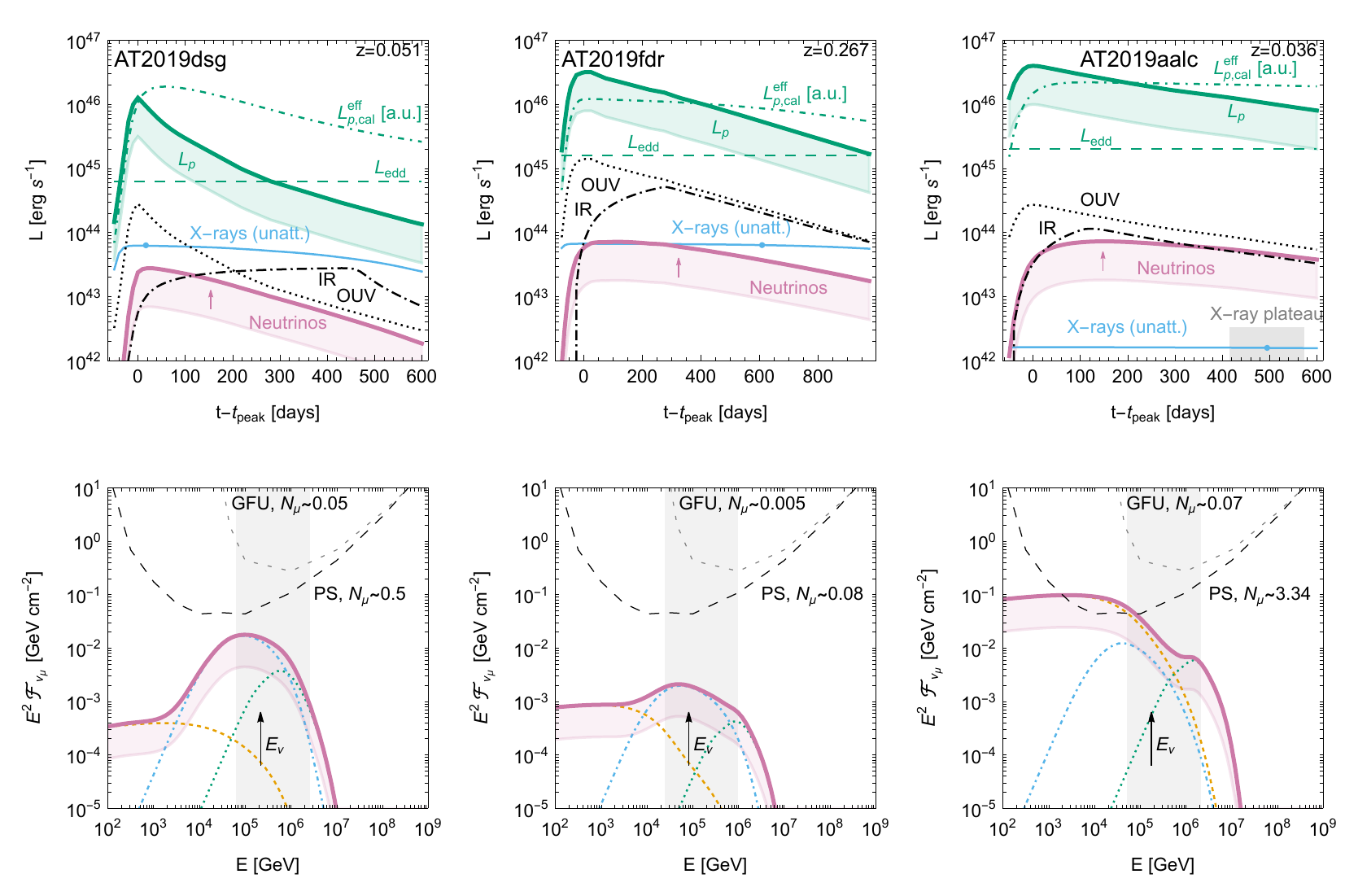} 
\end{center}
\caption{\label{fig:mx} Time-dependent evolution of the luminosities (upper row) and neutrino fluences (lower row) for model M-X.
In the upper row, our OUV (black dotted) and IR (black dashed-dotted) curves are shown as bolometric luminosities, OUV before dust attenuation. The X-ray curves show the assumed bolometric un-attenuated evolution from the disk, measurements are marked by dots. We also show the assumed proton injection luminosity $L_p$ and the in-source proton density as effective luminosity $L_{p,\mathrm{cal}}^{\mathrm{eff}}$ integrated over the highest  proton energies (arbitrary units, see main text for definition). The neutrino observation times are marked by arrows. In the lower row, the total neutrino fluence is shown, as well as its origin from different targets (outflow $pp$: dashed orange, X-rays: blue dashed-dotted, OUV: green dotted). In addition, the Gamma-Ray Follow-up (GFU)~\citep{Blaufuss:2019fgv} and Point-Source (PS)\citep{Aartsen:2014cva} differential limits  $E_\nu^2 \mathcal{F}_{\nu_\mu} =  E_\nu/(A_{\mathrm{eff}}(E_\nu) \, \mathrm{ ln } 10)$ are shown together with the integrated muon neutrino event rates (all events are in the same declination band). The mostly likely neutrino energies are marked, and gray-shaded areas mark the 99.73\% CL expected neutrino energy ranges expected for an $E^{-2}$ neutrino flux following the method in \citet{Palladino:2018evm}. Colored shadings correspond to varying $\varepsilon_{\mathrm{diss}}$ in the interval $\varepsilon_{\mathrm{diss}}=0.05 - 2$ (from lower to upper curves); $\varepsilon_{\mathrm{diss}}$ enters as multiplicative parameter in the proton injection and neutrino luminosities, see \equ{Lp}.}
\end{figure}

{\bf Results.}
Our main results for model M-X are presented in \figu{mx} for all three TDEs (columns). The upper row shows the time evolution of the luminosities (in the SMBH frame), the lower row reports the muon neutrino fluences and expected event rates, as well as the contributions from different targets. 
Flavor mixings are taken into account using the mixing angles in \citet{Esteban:2020cvm}. 
Concerning the time evolution (upper row), the proton injection luminosity follows the OUV BB luminosity (black dotted curves) in all cases. It scales proportionally to the Eddington luminosity (green dashed curves), so that at $t_{\mathrm{peak}}$ (by our assumptions) $L_p \simeq \varepsilon_{\mathrm{diss}} \, F_{\mathrm{peak}} \, L_{\mathrm{edd}} \simeq (5-20) \, L_{\mathrm{edd}}$ goes into non-thermal proton injection. The accumulated in-source proton density $N_p$ in \equ{kinetic} drives the neutrino production rate $J_\nu \propto N_p \, t^{-1}_{p \gamma}$ (with $t^{-1}_{p \gamma} \propto L_X$)  in the radiation zone (see \citet{Hummer:2010vx} for details); $N_p$ however cannot be directly shown in the figure because it is a density differential in energy and not in time; we consequently show a related effective luminosity integrated over the highest energies (chosen to be above the $p\gamma$ threshold, and therefore most relevant to \n\ production): 
\begin{equation}
L_{p,\mathrm{cal}}^{\mathrm{eff}}(t) \equiv t^{-1}_{\mathrm{fs}} \int\limits_{E_{p,\mathrm{max}}/3}^{E_{p,\mathrm{max}}} \, E_p \, N_p(E_p,t) \, dE_p \, ,
\label{eq:plcal}
\end{equation} 
see green dashed-dotted curves (in arbitrary units). We see that $L_{p,\mathrm{cal}}^{\mathrm{eff}}$ 
first increases with the proton injection, then it decays with a delay determined by $p\gamma$ interactions. The (unattenuated) X-ray luminosities are shown as blue curves, normalized to the highest {\bf observed} luminosity (blue dots). The neutrino light curves, at the leading order, then follow the product $L_{p,\mathrm{cal}}^{\mathrm{eff}} \times L_X$, but OUV and $pp$ interactions also contribute somewhat. Since $L_X$ is assumed to be roughly constant over $t_{\mathrm{dyn}}$, $L_{p,\mathrm{cal}}^{\mathrm{eff}}$ dominates the behavior of the neutrino light curve. The calorimetric behavior leads to a deviation between the solid ($L_p$) and dashed-dotted  ($L_{p,\mathrm{cal}}^{\mathrm{eff}}$) green curves here: if free-streaming escape dominated in \equ{kinetic}, the dashed-dotted curve would follow the solid curve  ($L_{p,\mathrm{cal}}^{\mathrm{eff}} \propto N_p \simeq J_p \, t_{\mathrm{fs}} \propto L_p$) -- whereas magnetically confined protons accumulate ($L_{p,\mathrm{cal}}^{\mathrm{eff}} \propto N_p \simeq \int J_p \, dt \propto \int L_p \, dt$) and lead to  $L_{p,\mathrm{cal}}^{\mathrm{eff}}$ being extended in time over a period comparable to $t_{\mathrm{dyn}}$. The observed neutrinos times are marked by arrows; the best description of the neutrino delay (peak consistent with arrow) is obtained for \aalc\, and the worst for \dsg\ (because of the quickly decaying BB). It is noteworthy that in all cases the neutrino luminosities (at the sources) are comparable, and the fluences at Earth are strongly affected by the redshifts. 

The lower row of \figu{mx} shows the neutrino fluences, which for \dsg\ and \fdr\ are dominated by X-ray interactions (blue dashed-dotted curves) with smaller contributions from the other targets. For \aalc , the observed X-ray luminosity was very low (see upper right panel), which means that indeed $pp$ interactions dominate here. The neutrino fluence is suppressed beyond about $20 \, \mathrm{TeV}$ by Bethe Heitler pair production off the OUV target photons. In the figure (as well as \figu{cartoon_MX}, right table) the neutrino event rates for the gamma-ray follow-up (GFU) and point source (PS) effective areas are shown for the respective declination band (that is similar for all TDEs).\footnote{The predicted event rates are computed by folding the predicted fluences with the corresponding effective areas over the full energy ranges: $N_\mu = \int \mathcal{F}_{\nu_\mu}(E_\nu) \, A_{\mathrm{eff}}(E_\nu) \, d E_\nu$. The GFU effective area includes the trigger probability of the gamma-ray follow-up pipeline, which implies that it has a higher energy threshold than the PS effective area. The GFU event rates should be used for a comparison to the actual observations from neutrino follow-ups, whereas the PS event rates are predictions relevant to evaluate if the source would appear in independent point source analyses, see discussion below and  in \Sec~\ref{sec:discussion}.   } A comparison of the GFU event number prediction with the expectation in \Tab~\ref{tab:observations} (for GFU, obtained in the listed references from counting statistics) indicates that (for both values of $\varepsilon_{\mathrm{diss}}$) the prediction for \dsg\ is in the expected range, that for \fdr\ is slightly below, and that for \aalc\, for which no comparison exists, is high: in fact, the source may appear in point source analyses. This indicates that $\varepsilon_{\mathrm{diss}} \gtrsim 0.1-0.2$ here, and it might be slightly different for the three TDEs. 

The predicted neutrino energies from X-ray interactions match observations very well (arrows), as expected; we note, however, that the probable neutrino energy has a range, illustrated by the gray areas in the figure (see figure caption for description), which extends to higher energies -- derived for an $E^{-2}$ input spectrum, though.

{\bf Discussion.}
Particle acceleration may occur in high-velocity winds embedded in the TDE debris  \citep{Murase:2020lnu} or shocks from stream crossings \citep{Hayasaki:2019kjy}. Compared to \citet{Murase:2020lnu}, our production region is typically slightly smaller (than $10^{16}$~cm) and our required cosmic-ray injection luminosity (which was normalized to the BB luminosity in that paper) is higher, as it can be seen in \figu{mx} (upper row). This means that in the hidden wind models it may be difficult to achieve high enough proton luminosities (i.e., $\varepsilon_{\mathrm{diss}}$)  at least from the estimates in \citet{Murase:2020lnu}. Magnetic confinement is also considered in the hidden wind model, where however adiabatic losses limit the optical thickness.
  In the hidden wind models, $pp$ interactions with the debris can also play a major role; however large uncertainties are implied, such as from the geometry and the time evolution of the system.
 There are also some similarities with the TDE outflow model for \dsg\ in \citet{Wu:2021vaw}: outflow-cloud interactions may lead to particle acceleration, and $pp$ interactions with the clouds may lead to neutrino production. While the production region in that model is a bit larger ($R \simeq 10^{16} \, \mathrm{cm}$), the $pp$ interactions are efficient in the clouds, which are assumed to have a size of about $\simeq 10^{14} \, \mathrm{cm}$ and act as calorimeters. Our $pp$ interactions with the outflow itself are less efficient because our production radius is large (the target density is about a factor 50 smaller), and therefore the effect is sub-dominant for \dsg\, see \equ{tppfs}. We note however that the shocks generated by the outflow-cloud interactions may be an interesting acceleration sites. For a comparison/discussion of jetted models, see \App~\ref{app:jetted}. Furthermore, if the radiation zone is expanding -- such as expected in the wind models -- adiabatic cooling can affect especially model M-X, see \App~\ref{app:adiabatic}; we do not consider this effect in the main text because it does not apply to accelerators with a stable radiation zone, and we anticipate that winds or outflows as accelerators are furthermore challenged by the high required $\varepsilon_{\mathrm{diss}}$. Off-axis jets may be an alternative especially since the calorimeter leads to the isotropization of protons emitted into different directions; here the emission radius of the cosmic rays would need to match $R$ -- which is quantitatively comparable to the collision radius expected for internal shocks~\citep{Lunardini:2016xwi}. The potential contributions from core models cannot be captured by our approach because of the assumption $R \gtrsim R_{\mathrm{BB}}$, which means that accelerators in the core (where interactions are more efficient) cannot be well described for model M-X.

Finally we note that in all cases an average neutrino delay $t_\nu-t_{\mathrm{peak}}$ in the right ball park is expected, originating in the calorimetric behavior of the system. However, the neutrino light curves are widely spread in time, and a slight preference of an early neutrino emission closer to the peak is implied in the cases with higher $p \gamma$ efficiencies -- which have the effect of depleting the available in-source protons. Thus a high neutrino production efficiency and a long neutrino delay are  anti-correlated for model M-X. An exception is \aalc\, where (slow) $pp$ interactions dominate and the neutrino flux peaks at lower energies. Ways to improve the neutrino time dependence emerge if the non-thermal proton injection is delayed with respect to the mass fallback rate. Note that in all cases the actual X-ray emission may be even higher than the observation, since the observed flux may be obscured as well. However, this may be unlikely for \aalc\, where the observed X-ray flux was stable over the duration of a few hundred days (called ``X-ray plateau'' in figures).  

\section{Medium-energy protons interacting with optical-UV photons (model M-OUV)}
\label{sec:mouv}

\begin{figure}[t]
    \centering
    \begin{tabular}{cc}
    \includegraphics[width=0.5\textwidth]{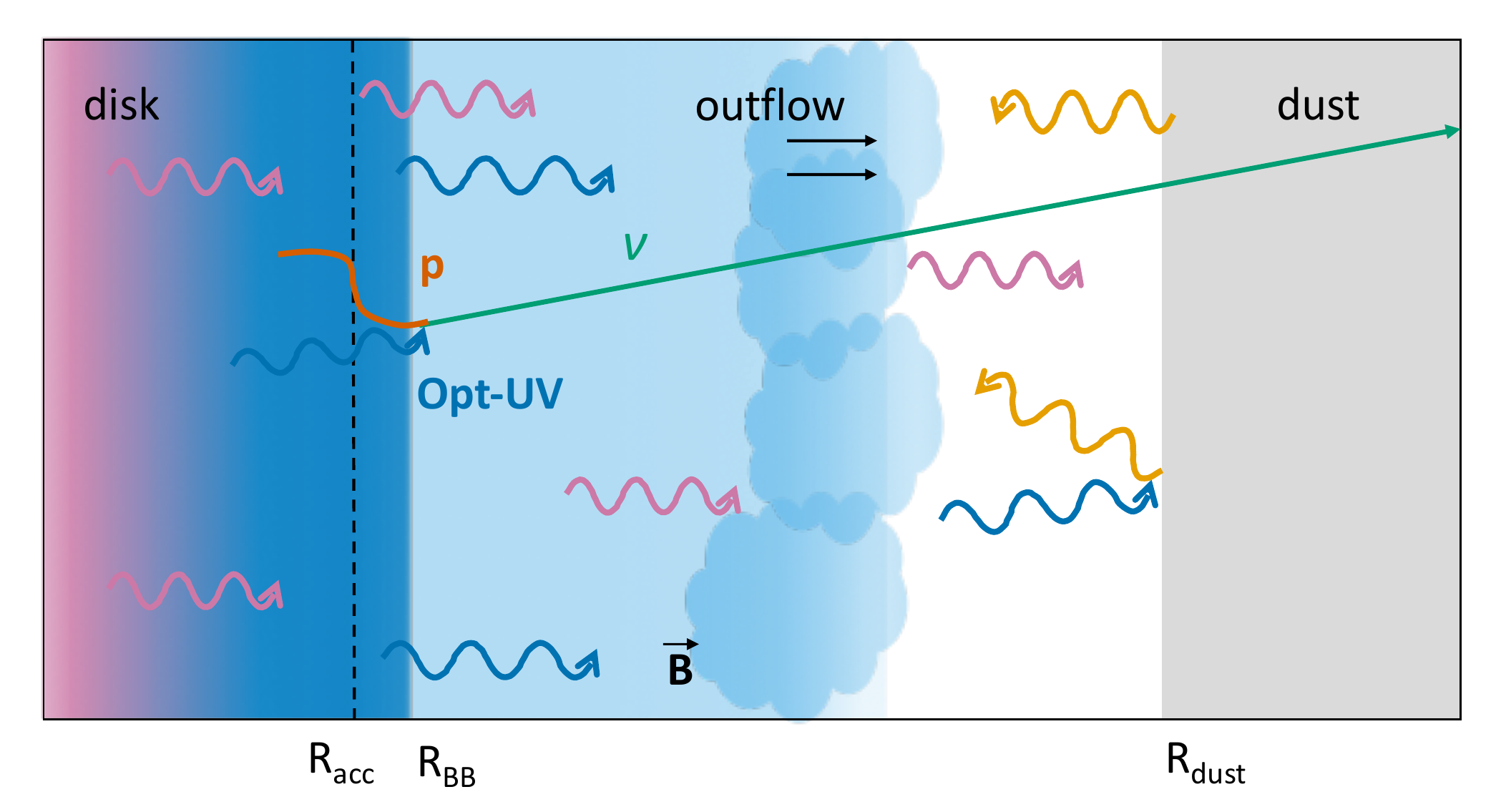}  &  \hspace*{-2cm}
     \raisebox{2.3cm}{
     \begin{tabular}{|l|ccc|}
    \hline
    & {\bf \dsg} & {\bf \fdr} & {\bf \aalc} \\
       \hline
     \multicolumn{4}{|l|}{{\bf Model assumptions}} \\
     \hline
        $R$ [cm] & $5.0 \, 10^{14}$ &  $4.9 \, 10^{15}$ &  $1.0 \, 10^{16}$ \\
        $E_{p,\mathrm{max}}$ [GeV]  & $1.0 \, 10^8$  &
        $1.0 \, 10^8$  &   
        $1.0 \, 10^8$ \\
        $B$ [G] & 0.1 &  0.1 & 0.1   \\
        \hline
     \multicolumn{4}{|l|}{{\bf Model results}} \\
     \hline
     $N_\nu$ GFU  & 0.21  & 0.055  &  1.2   \\
     & (0.052) & (0.014) & (0.29) \\
     $N_\nu$ PS  & 1.1  & 0.22   & 3.6   \\
     & (0.28) & (0.054) & (0.90) \\
     \hline
    \end{tabular}
    } 
    \end{tabular}
    \caption{Model M-OUV. Left panel: Microphysics cartoon. Right panel: Model assumptions and results. Here $R \sim R_{\mathrm{BB}}$ from \citet{vanVelzen:2020cwu} (\dsg) and from \citet{Reusch:2021ztx} (\fdr), whereas it is an assumption for \aalc . Predicted event rates are for $\varepsilon_{\mathrm{diss}}=0.2$ (0.05). }
    \label{fig:cartoon_MOUV}
\end{figure}

{\bf Model-specific description.}
In order to foster interactions with the OUV target photons, $E_{p,\mathrm{max}} = 1 \cdot 10^8 \, \mathrm{GeV}$, higher than for M-X, is used here. The microphysics is illustrated in the cartoon in \figu{cartoon_MOUV}, left panel, and model assumptions and results are listed in the table in \figu{cartoon_MOUV}, right panel. If $R_{\mathrm{acc}} \lesssim R_{\mathrm{BB}}$, the radiation zone is determined by the black body radius  $R \simeq  R_{\mathrm{BB}}$ for M-OUV, for which we use measured or estimated values listed in the table. While possible accelerators could (in principle) be disk or corona, we assume for the sake of simplicity of computation that $R_{\mathrm{acc}} \sim R \sim  R_{\mathrm{BB}}$.\footnote{This implies that, for the sake of generality, we neglect additional interactions with X-rays in the core, which are present for the core models (small $R_{\mathrm{acc}}$):
For X-rays, the optical thickness increases with decreasing injection radius $R_{\mathrm{acc}}$ as long as  $R_{\mathrm{acc}} > R_X$ (beyond the X-ray photosphere). However, it can be estimated that in the core $\tau_{p \gamma}^{\mathrm{fs}}$, describing the X-ray interactions in \equ{taumx},  is for $R \gtrsim R_S$ (for $M=10^7 \, M_\odot$) smaller than $\tau_{p \gamma}^{\mathrm{fs}}$ in \equ{tpgfsOUV}, describing the OUV interactions for $R\simeq R_{\mathrm{BB}}$. This means that the additional core contributions are smaller than the ones off the OUV target for the chosen parameters and comparable proton injection luminosities. }

While the OUV luminosity has been measured for all three TDEs, a dust echo has been measured as well in each case. Its intensity implies that the actual BB luminosity at $R_{\mathrm{BB}}$ must be significantly higher than the observed luminosity; we therefore derived in \Sec~\ref{sec:dust} an estimate for the minimal bolometric luminosity, see \Tab~\ref{tab:observations}. Furthermore, the target photon density in the BB photosphere is estimated by the free-streaming assumption \equ{norm}, which is conservative if $R_{\mathrm{acc}} < R_{\mathrm{BB}} \simeq R$, and more accurate if $R_{\mathrm{acc}} \simeq R \gtrsim R_{\mathrm{BB}}$. Therefore, we anticipate that our neutrino prediction for M-OUV is on the conservative side, and the actual neutrino fluence could be somewhat higher.

\begin{figure}[t]
\begin{center}
\includegraphics[width=0.45\textwidth]{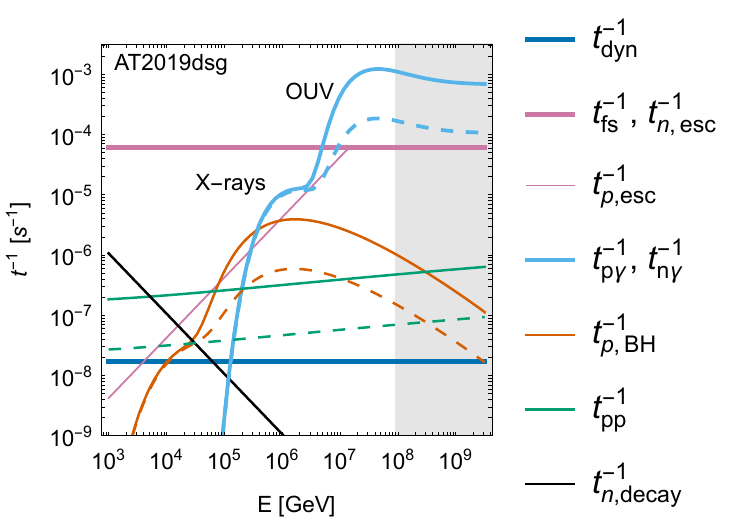} 
\end{center}
\caption{\label{fig:timescales_MOUV} Relevant inverse timescales (rates) for protons and neutrons for M-OUV and \dsg\ in the SMBH frame as a function of observed energy. For details, see caption of \figu{timescales_MX}. }
\end{figure}

{\bf Interaction rates and optically thick behavior.}
In \figu{timescales_MOUV}, we show the rates (inverse timescales) as a function of energy for \dsg\ as an example, for the parameters chosen for M-OUV; again solid curves refer to peak time and dashed curves to the time of the neutrino emission. Here $E_{p,\mathrm{max}}$ is high enough that interactions with the OUV target are possible -- and, in fact, they dominate even at the time of the neutrino emission. The optical thickness to $p\gamma$ interaction can be analytically estimated as
\begin{equation}
 \tau_{p \gamma}^{\mathrm{fs}} \equiv \frac{t_{\mathrm{fs}}}{t_{p \gamma}}  \simeq 300 \, \left( \frac{L_{\mathrm{BB}}}{10^{45} \, \mathrm{erg \, s^{-1}}} \right) \left( \frac{T_{\mathrm{BB}}}{\mathrm{eV}} \right)^{-1} \left( \frac{R}{10^{15} \, \mathrm{cm}} \right)^{-1}\, ,
\label{equ:tpgfsOUV}
\end{equation}
which means that typically  $\tau_{p \gamma}^{\mathrm{fs}} \gg 1$ at peak, and the system is optically thick and
 protons and neutrons cool efficiently by $p\gamma$ interactions.

X-ray and $pp$ interactions may still contribute at lower energies over the dynamical timescale, as the $p \gamma$ interaction rate off X-rays is higher than the (diffusive) escape rate. While the OUV interaction time is of the order of hours, the neutrinos from X-ray and $pp$ interactions tend to come later. Nevertheless, it is clear already from the interaction rates that the dominant contribution to the neutrino signal (from OUV interactions) will follow the BB evolution on the hour scale. 

\begin{figure}[t]
\begin{center}
\includegraphics[width=\textwidth]{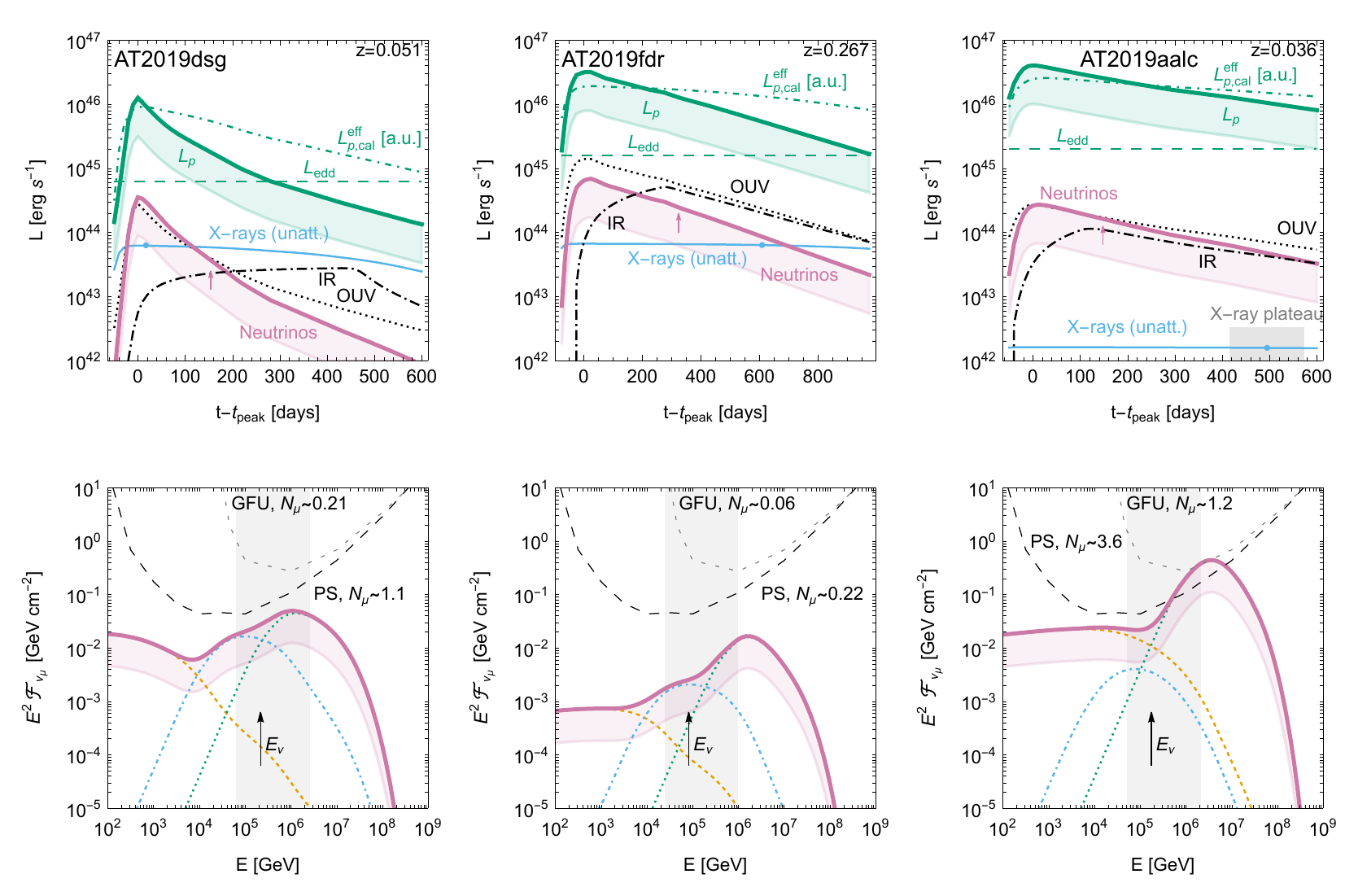} 
\end{center}
\caption{\label{fig:ouv} Time-dependent evolution of the luminosities (upper row) and neutrino fluences (lower row) for model M-OUV. For details, see caption of \figu{mx}.}
\end{figure}

{\bf Results.} \figu{ouv} displays our main results in the same format as \figu{mx}. Here the neutrino light curves (reddish-purple curves in upper panels) follow the product of $L_{p,\mathrm{cal}}^{\mathrm{eff}}$ (green dashed-dotted curve) and $L_{\mathrm{BB}}^{\mathrm{bol}}$ (min.) (black dotted curve) at the leading order. As expected, it is more difficult to reproduce the neutrino time delay marked by the reddish-purple arrows.
Here $L_{p,\mathrm{cal}}^{\mathrm{eff}} \propto N_p \simeq J_p \, t_{p\gamma} \propto L_p \, L_{\mathrm{BB}}^{-1}$ from \equ{kinetic}. The fact that both $L_p$ and $L_{\mathrm{BB}}$ have the same time dependence by construction ($L_p$ follows the BB light curve, and the interaction rate depends on that as well) explains why $L_{p,\mathrm{cal}}^{\mathrm{eff}}$ declines over time more slowly than $L_p$, until at later times escape takes over as leading radiation mechanism. It is not a calorimetric behavior here, but a consequence of the explicit time-dependence of the injection and target luminosities. The neutrino luminosity is $\propto L_{p,\mathrm{cal}}^{\mathrm{eff}} \times L_{\mathrm{BB}}$, and therefore it, to first approximation, follows the strong time-dependence of $L_{\mathrm{BB}}$. 

The neutrino fluences in the lower row (see also \figu{cartoon_MOUV}, right table) are higher than for model M-X, and are all within the expected ranges in \Tab~\ref{tab:observations}. The peaks are all dominated by the BB interactions (green dotted curves) thanks to the high enough proton energies, whereas X-rays (blue dashed-dotted curves) and $pp$ (orange dashed curves) interactions contribute at lower energies. The predicted neutrino energies are significantly higher than the most likely energies (black arrows), but are plausible if the uncertainties (gray areas) are taken into account. It is noteworthy that \aalc\  exhibits event rates $N_\mu > 1$ for both the GFU and PS effective areas -- if $\varepsilon_{\mathrm{diss}}$ is largest, $\varepsilon_{\mathrm{diss}}=0.2$ --  in spite of the relatively large assumed value for $R$. This means that, while the observation of \dsg\ and \fdr\ may have been a coincidence motivated by the Eddington bias, \aalc\ is a neutrino source which could have been detected independently as neutrino point source.

{\bf Discussion.}
While M-OUV exhibits higher event rates than M-X, it offers a poorer description of the neutrino time delay, and the size of the system plays only a minor role in this. A possible solution of this problem might be that the neutrino delay comes from a delayed proton injection itself, for which we could however not yet identify a physical motivation.
The strength of model M-OUV is that high neutrino fluences can be produced, given that the estimates for the unobscured BB luminosities are lower limits only. Comparing the expected number of neutrinos in \Tab~\ref{tab:observations} with the predicted ones in \Fig~\ref{fig:cartoon_MOUV} (right table), one finds that $\varepsilon_{\mathrm{diss}} \gtrsim 0.03$ satisfies all requirements. 

As pointed out earlier, our model M-OUV is a quantitative implementation of the original analytical estimate in \citet{Stein:2020xhk} for \dsg ; therefore, we have performed numerical cross-checks for similar parameters to isolate the differences.  From \figu{ouv}, upper left panel, we can see that the neutrino luminosity is about $L_p/40$  at $t_{\mathrm{peak}}$, whereas the original model predicts a factor $L_p/8$ in the optically thick case. The main difference comes from the bolometric correction: our $L_p$ is related to the full non-thermal proton spectrum, not only to the part beyond the pion production threshold; a smaller contribution comes from the pitch-angle averaging and width of the cross section taken into account in numerical computations. Furthermore, note that the optical thickness drops as a function of time, which overall leads to a significantly lower neutrino event rate prediction than the analytical estimate taken at the BB peak.

Let us discuss possible accelerator realizations. First of all, note that if an explicit acceleration rate is considered, the $E_{p,\mathrm{max}}$ in \figu{timescales_MOUV} cannot be self-consistently reached if the radiation and acceleration zones are identical, see discussion in \App~\ref{app:maxproton}. Therefore, a different, possibly more compact acceleration region below the OUV photosphere with stronger magnetic fields is implied, which might be a disk or corona; however, note that high enough proton energies must be reached, which seems challenging for stochastic acceleration~\citep{Murase:2019vdl}. An off-axis jet is unlikely here, because the cosmic rays would interact faster than they can isotropize, which means that the neutrinos would propagate in the jet direction (see \App~\ref{app:jetted}), whereas wind or outflow models are disfavored for M-OUV because they typically require $R \gg R_{\mathrm{BB}}$. 

\section{Ultra-high-energy protons interacting with infrared photons (model M-IR)}
\label{sec:mir}

{\bf Model-specific description.}
The fact that all three neutrino-observed TDEs have been associated with strong dust echoes suggests a direct connection of the neutrino production with the IR photons from the dust echoes. Even more: the dust echo light curves, shown as dashed-dotted black curves in our results plots (\eg\ \figu{mx}), seem to directly suggest being the origin of the neutrino time delay (see arrows for neutrino arrivals), because the neutrinos arrive at or close to the peak of the dust echo.\footnote{For \dsg , the dust echo peaks much later, but the OUV light curve much earlier, which may compensate the delay.}

\begin{figure}[t]
    \centering
    \begin{tabular}{cc}
    \includegraphics[width=0.5\textwidth]{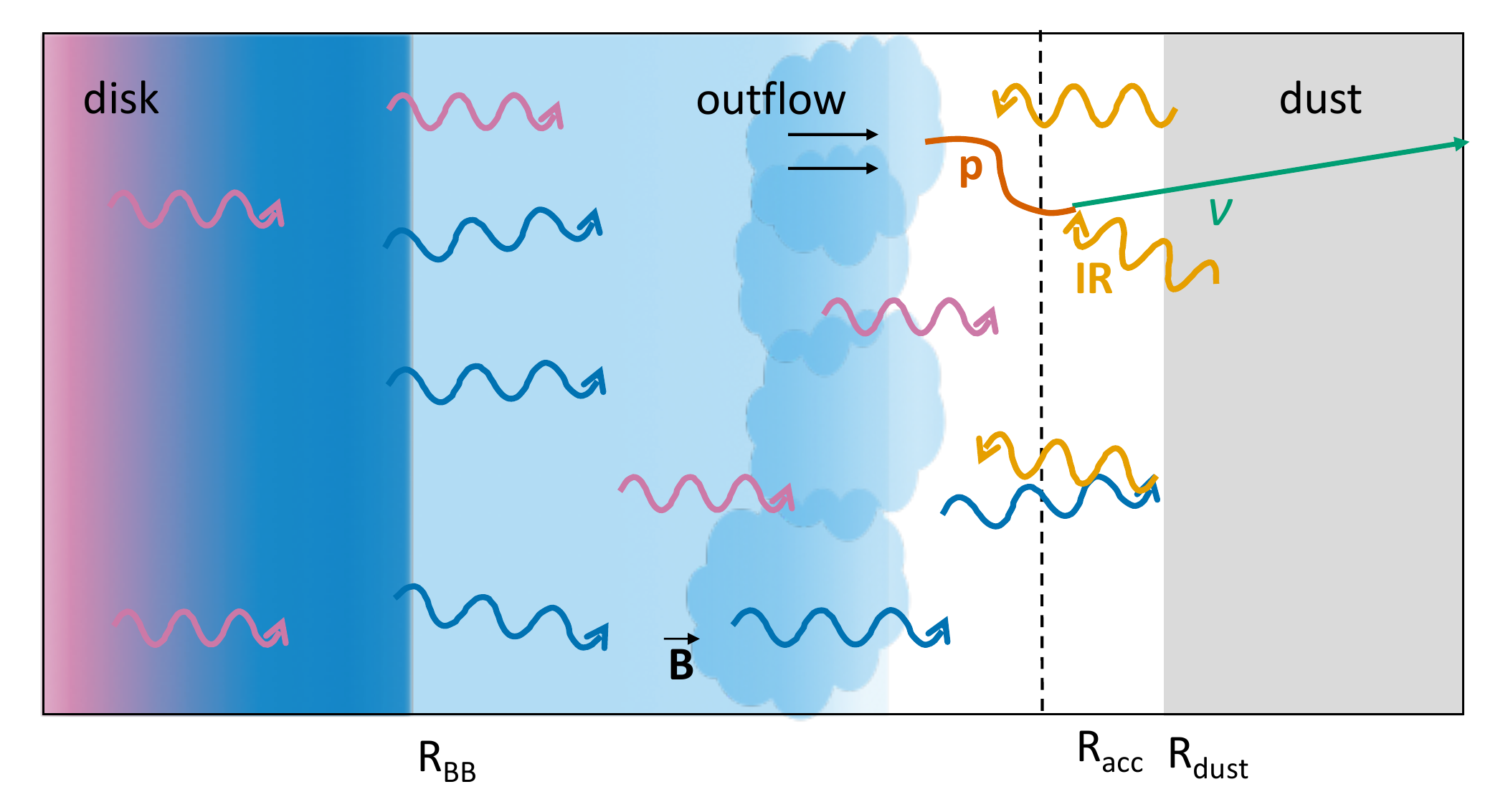}  & \hspace*{-2cm}
     \raisebox{2.5cm}{
      \begin{tabular}{|l|ccc|}
    \hline
     & {\bf \dsg } & {\bf \fdr } & {\bf \aalc } \\
       \hline
     \multicolumn{4}{|l|}{{\bf Model assumptions}} \\
     \hline
        $R$ [cm] & $5.0 \, 10^{16}$ &  $2.5 \, 10^{17}$ &  $2.0 \, 10^{17}$ \\
        $E_{p,\mathrm{max}}$ [GeV]  & $5.0 \, 10^9$  &
        $5.0 \, 10^9$  &   
        $5.0 \, 10^9$ \\
        $B$ [G] & 0.1 &  0.1 & 0.1   \\
        \hline
     \multicolumn{4}{|l|}{{\bf Model results}} \\
     \hline
     $N_\nu$ GFU  & 0.030  & 0.013  &  0.036   \\
      & (0.0074) & (0.0032) & (0.091) \\
     $N_\nu$ PS  & 0.074  & 0.024   & 0.62   \\
     & (0.019) & (0.0060) & (0.16) \\
     \hline
    \end{tabular}   
    } 
   \end{tabular}
    \caption{Model M-IR. Left panel: Microphysics cartoon. Right panel: Model assumptions and results. Here  $R$ is taken from \citet{Stein:2020xhk} (radio), \citet{Reusch:2021ztx} (IR) and \Tab~\ref{tab:observations} ($c \, \Delta t$) for \dsg , \fdr , and \aalc, respectively. Predicted event rates are for $\varepsilon_{\mathrm{diss}}=0.2$ (0.05).  }
    \label{fig:cartoon_MIR}
\end{figure}

The microphysics of model M-IR is illustrated in the cartoon in \figu{cartoon_MIR}, left panel, and model assumptions and results are listed in the table in \figu{cartoon_MIR}, right panel. The challenge here are the ultra-high required proton and corresponding neutrino energies; we use a TDE-universal $E_{p,\mathrm{max}} = 5 \, 10^9 \, \mathrm{GeV}$, see \equ{target} and discussion thereafter. The scattering and re-processing of the OUV (perhaps even X-ray) photons in the dust leads to a delayed IR signal to the observer, as outlined in \Sec~\ref{sec:dust}; isotropized IR photons will correspondingly fill the production volume with  $R \simeq R_{\mathrm{dust}}$. The inferred $R_{\mathrm{dust}}$ of the order of $10^{17} \, \mathrm{cm}$ is typically a rough estimate from the time delay of the dust echo, subject to geometric uncertainties. We choose the values in \figu{cartoon_MIR}, right table, which are related to IR observations (\fdr) or our dust model (\aalc). For \dsg , we note that too large values of $R$ lead to too inefficient neutrino production; for our dust model, the observed $\Delta t$ in \Tab~\ref{tab:observations} translates into $R_{\mathrm{dust}} \simeq 6 \, 10^{17} \, \mathrm{cm}$, i.e., the largest dust region --  compared to the smallest $R_{\mathrm{BB}}$. Instead of this value, we use the size $R$ inferred from the observed radio signal, speculating that dust region may be more structured to satisfy all constraints. The chosen value for $B=0.1 \, \mathrm{G}$ for all three TDEs leads to confinement of protons over such a large region, see \equ{dx}, which helps the reproduction of the neutrino time delay.
The IR luminosities and light curves are taken from our own dust model in \Sec~\ref{sec:dust}, the in-source density is computed with \equ{norm}.

\begin{figure}[t]
\begin{center}
\includegraphics[width=0.45\textwidth]{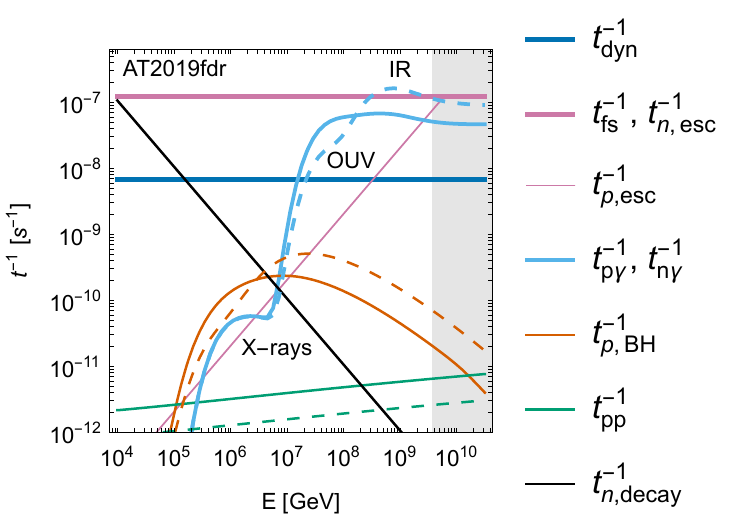} 
\end{center}
\caption{\label{fig:timescales_MIR} Relevant inverse timescales (rates) for protons and neutrons for M-IR and \fdr\ in the SMBH frame as a function of observed energy. For details, see caption of \figu{timescales_MX}. }
\end{figure}

{\bf Interaction rates and optically thick/calorimetric behavior.}
In \figu{timescales_MIR}, we show the rates (inverse timescales) as a function of energy this time for \fdr\ as an example, for the parameters chosen for M-IR; again solid curves refer to peak time and dashed curves to the time of the neutrino emission. Here $E_{p,\mathrm{max}}$ is high enough that interactions with all targets are possible. We see that the free-streaming and dynamical timescales are much closer to each other than in the other models because of the larger size of the region; interactions with X-rays and with the outflow (pp) play a minor role here, so does Bethe Heitler pair production. However, the interactions with the OUV photons, which have a higher luminosity but not necessarily number density, cannot be neglected. At $t_{\mathrm{peak}}$, the system is optically thin to $p\gamma$ interactions; however, confined protons will interact over $t_{\mathrm{dyn}}$ with the OUV and IR targets similarly to M-X. At $t_{\nu}$, the interactions with IR photons from the dust echo in fact dominate the neutrino production. 

The analytical estimates for the  optical thicknesses in the free-streaming and calorimetric cases and the IR target are 
\begin{eqnarray}
\tau_{p \gamma}^{\mathrm{fs}} &&\equiv \frac{t_{\mathrm{fs}}}{t_{p \gamma}} \simeq 3 \,
\left( \frac{L_{\text{IR}}}{10^{44} \, \mathrm{erg \, s^{-1}}} \right) 
\left( \frac{T_{\text{IR}}}{0.1 \, \mathrm{eV}} \right)^{-1} \left( \frac{R}{10^{17} \, \mathrm{cm}} \right)^{-1}~, 
\nonumber \\
 \tau_{p \gamma}^{\mathrm{cal}} &&\equiv \frac{t_{\mathrm{dyn}}}{t_{p \gamma}} \simeq 60 \,
\left( \frac{L_{\text{IR}}}{10^{44} \, \mathrm{erg \, s^{-1}}} \right) 
\left( \frac{T_{\text{IR}}}{0.1 \, \mathrm{eV}} \right)^{-1} \left( \frac{R}{10^{17} \, \mathrm{cm}} \right)^{-2}
\left( \frac{t_{\mathrm{dyn}}}{900 \, \mathrm{days}} \right)~,
\label{equ:tpgIR}
\end{eqnarray}
respectively. This means that the neutrino production from protons interacting with the IR target is guaranteed to be efficient over the dynamical timescale, and that the system may be optically thick.  Assuming $t_{p \gamma,\mathrm{cool}} \simeq 5 \times t_{p \gamma}$ (such as in the optically thick case), it is also interesting that the proton interactions are slow: 
\begin{equation}
t_{p \gamma,\mathrm{cool}}  \simeq 70 \,\mathrm{days} \left( \frac{T_{\text{IR}}}{0.1 \, \mathrm{eV}} \right) \left( \frac{L_{\text{IR}} }{10^{44} \, \mathrm{erg \, s^{-1}}} \right)^{-1} \left( \frac{R}{10^{17} \, \mathrm{cm}} \right)^{2}\, , 
\label{equ:delay}
\end{equation}
which helps the neutrino time delay, as the injected in-source protons will be depleted over that time scale.

\begin{figure}[t]
\begin{center}
\includegraphics[width=\textwidth]{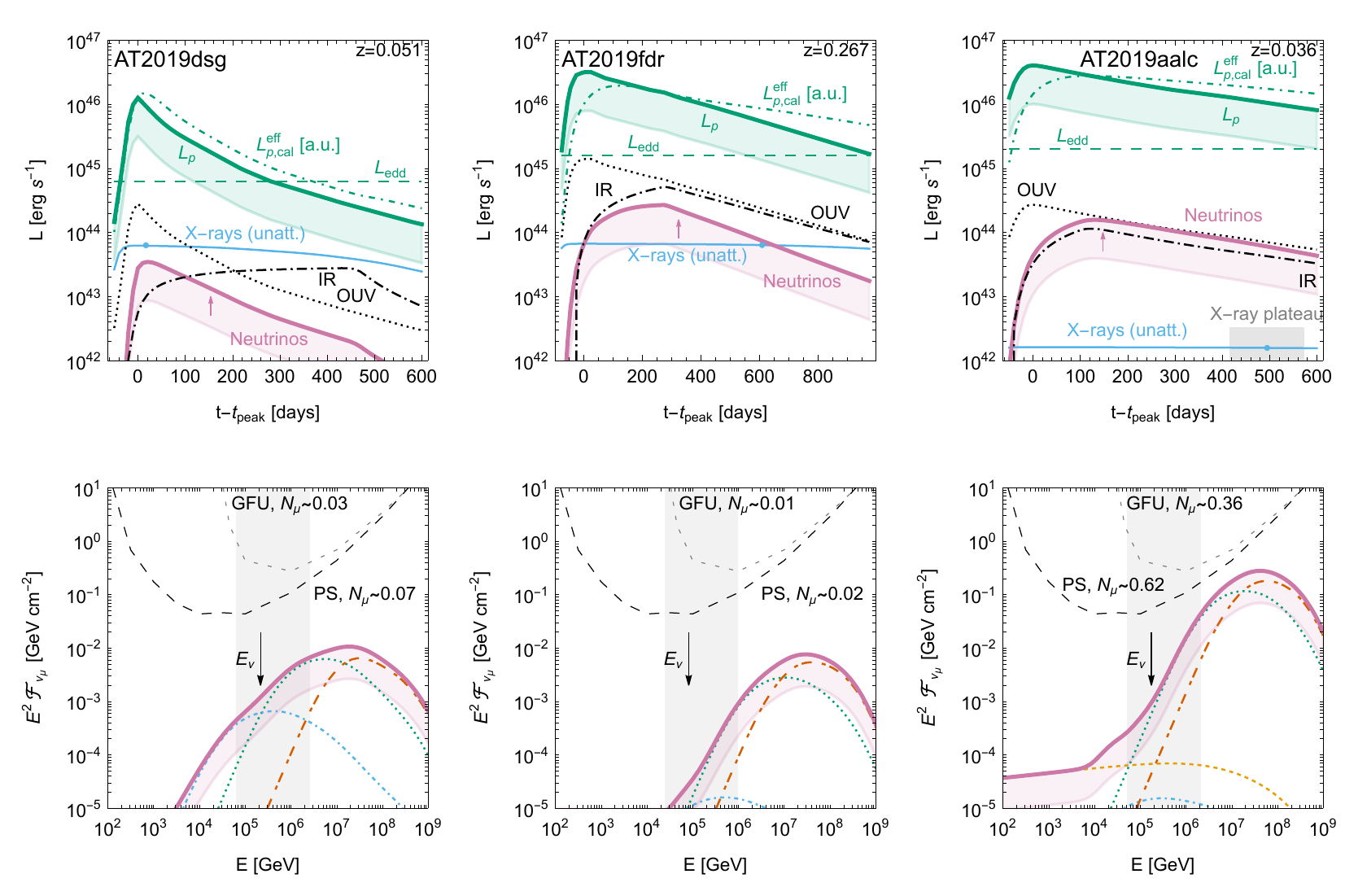} 
\end{center}
\caption{\label{fig:ir} Time-dependent evolution of the luminosities (upper row) and neutrino fluences (lower row) for model M-IR. For details, see caption of \figu{mx}, the additional contribution from the IR photon interactions is red dashed-dotted in the lower row.}
\end{figure}

{\bf Results.}
We present our main results for model M-IR in \figu{ir}. 
The neutrino light curves (reddish-purple curves in upper panels) follow the product of $L_{p,\mathrm{cal}}^{\mathrm{eff}}$ (green dashed-dotted curves) and OUV (black dotted curves) or IR (black dashed-dotted curves). Therefore, a different time evolution for the neutrinos from OUV and IR is expected. Here we re-discover a calorimetric behavior similar to M-X for \fdr\ and \aalc\, in which cases the neutrino time delay (reddish-purple arrows) can be easily reproduced from the time delay of the dust echo. On the other hand, \dsg\ is free streaming escape-dominated at the highest energies, which means that $L_{p,\mathrm{cal}}^{\mathrm{eff}} \propto N_p \simeq J_p \, t_{\mathrm{fs}} \propto L_p$ from \equ{kinetic}; the reason is the more compact production region leading to higher escape rates.
Since the target photon density is relatively constant over a wide time window and the neutrino luminosity is $\propto L_{p,\mathrm{cal}}^{\mathrm{eff}} \times L_{\mathrm{IR}}$, it follows $L_p$ to leading order for this TDE. 

The neutrino fluences in the lower row (see also \figu{cartoon_MIR}, right table) are more comparable to M-X rather than to M-OUV. The predicted numbers of events lie  within the expected ranges in \Tab~\ref{tab:observations} for $\varepsilon_{\mathrm{diss}} \gtrsim 0.1$. The peaks of the fluences are all dominated by the IR interactions (orange dashed-dotted curves) because of the high enough proton energies, but OUV interactions (blue dotted curves) contribute significantly in all cases. This is unavoidable, since IR and OUV are directly connected, which means that a part of the OUV luminosity is re-processed into the IR dust echo -- and consequently the OUV interactions (which also have a lower threshold) cannot be avoided.   The predicted neutrino energies are significantly higher than the most likely energies (black arrows), which is the biggest limitation of model M-IR. 

{\bf Discussion.}
Model M-IR provides the best description of the neutrino time delay due to the time evolution of the dust echo target, with the exception of \dsg . The neutrino delay description could be improved somewhat by a larger value of $R$ (because the protons interact more slowly), at the expense of the overall neutrino production efficiency. Therefore there is no good tradeoff between time delay and neutrino fluence for \dsg .  Major disadvantages of model M-IR are the predicted neutrino energies peaking at very high values (which are significantly above the detected energies), and the relatively low predicted neutrino fluences.  

The location of the accelerator $R_{\mathrm{acc}}<R_{\mathrm{dust}} \simeq R$ can be, in principle, anywhere in the isotropization region. However, we know from the discussion of  \equ{larmor} that the confinement condition already indicates that $R_{\mathrm{acc}} \gtrsim 3 \, 10^{16} \, \mathrm{cm}$ for $E_{p,\mathrm{max}} \simeq \mathrm{EeV}$ and $B=0.1 \, \mathrm{G}$, which points towards $R_{\mathrm{acc}} \sim R$; a wind or outflow could serve as accelerator, which is however challenged by the required large $\varepsilon_{\mathrm{diss}}$. The acceleration may also occur in more compact regions in the core (corona, disk) or a jet with larger values of $B$; especially in the case of the core models, additional contributions would add to the neutrino signal from the more compact regions, which we do not describe here. 

We would like to note that the model M-IR could be connected with origin of the UHECRs (which is why we refer to ``ultra-high'' proton energies in this section). Depending on nuclear disintegration and air shower models, a maximum rigidity  $R_{\mathrm{max}} \simeq (1.4 - 3.5) \, 10^9 \, \mathrm{GV}$ was found in \citet{Heinze:2019jou} to describe data from the Pierre Auger Observatory~\citep{PierreAuger:2016use} in the rigidity-dependent maximal energy model in multi-dimensional parameter space fit. This range  is sufficiently close to the assumption made for protons here  $E_{p,\mathrm{max}} = 5 \, 10^9 \, \mathrm{GeV}$, which means that TDEs similar to the ones discussed in this work could also be the sources of the UHECR protons. For heavier nuclear compositions, needed to describe UHECR data, the neutrino fluence is expected to be similar for an $E^{-2}$ cosmic-ray acceleration spectrum (see \citet{Morejon:2019pfu} for the dependence of $A\gamma$ interactions on the mass number) as long as the source is optically thin to nuclear disintegration at the highest energies, see \citet{Biehl:2017hnb,Morejon:2019pfu} for corresponding examples. Of course, the nuclear composition might be motivated by appropriate progenitor disruptions, such as ONe or CO white dwarfs. Concerning the energy output in UHECRs, we know that the energy in non-thermal protons per TDE is $E_p \simeq \varepsilon_{\mathrm{diss}} \int \dot M dt \simeq 0.1 \, M_\star$ in our model for the optimistic $\varepsilon_{\mathrm{diss}}=0.2$, where $M_\star$ is given in \Tab~\ref{tab:observations}. Taking into account the bolometric correction (only protons at the highest energies are relevant here) and that only a fraction of UHECR protons escape (see \eg~\citet{Heinze:2020zqb} for a corresponding discussion in GRBs), one may estimate that $E_p^{\mathrm{esc}} \lesssim 0.01 \, M_\star$ can be re-processed into non-thermal protons at the highest energies per TDE; for a solar-mass star, that is about $2 \, 10^{52} \, \mathrm{erg}$. This would require a local white dwarf disruption rate of about $5 \, \mathrm{Gpc}^{-3} \, \mathrm{yr}^{-1}$ to match a local injection rate of $10^{44} \, \mathrm{erg} \, \mathrm{Mpc}^{-3} \, \mathrm{yr}^{-1}$ -- which has been perceived as too high in the literature, see \eg\ discussion in \citet{Biehl:2017hnb}. While more recent observations find much higher local rates,  $500 \, \mathrm{Gpc}^{-3} \, \mathrm{yr}^{-1}$~\citep{Tanikawa:2021zfm}, it is critical here how the population extends to large SMBH masses (as the UHECR output in our model scales with $M \propto L_{\mathrm{edd}}$); a detailed population model is beyond the scope of this study. Another possibility could be that material from the pre-existing AGNs is injected into the acceleration process, which would impact both energy budget and UHECR composition.

\section{Predicted diffuse neutrino fluxes}
\label{sec:diffuse}

Using the results in the previous sections, we have computed the expected diffuse flux of \ns\ from \tds, following the method outlined in \citet{Lunardini:2016xwi}. The \n\ flux of a given flavor $\alpha$ -- differential in energy, time, area and solid angle --  is given by: 
\begin{equation}
\Phi_\alpha(E) =\frac{\eta \, c}{4 \pi H_0} \int_{M_{\rm min}}^{M_{\rm max} } dM  \int_{0}^{z_{\rm max}}  dz\ \frac{ \dot{\rho}(z,M) Q_\alpha(E(1+z),M)}{\sqrt{\Omega_M (1+z)^3+\Omega_\Lambda}}~,
\label{equ:diffmaster}
\end{equation}
where $\dot{\rho}(z,M)$ is the cosmological rate of \tds\ differential in redshift and \bh\ mass (obtained following \citet{Shankar:2007zg,Stone:2014wxa,Kochanek:2016zzg}; see also  \citet{Holoien:2015pza,vanVelzen:2017qum,Hung:2017lxm} for rate measurements). 
The function $Q_\alpha$ is the number of neutrinos emitted per unit energy in the \bh\ frame (inclusive of neutrino flavor oscillations in vacuum) for a single \td\ having a \bh\ of mass $M$, taken from our results of Secs. \ref{sec:mx}, \ref{sec:mouv} and \ref{sec:mir} respectively, for each of the three models; it corresponds to the quantity $J$ in \equ{kinetic} integrated over the volume of the radiation zone and over the emission time.
 The quantity $E^\prime=E(1+z)$ is the \n\ energy in the SMBH frame; $E$ is the energy observed at Earth. Here $\eta$ is the fraction of \tds\ where the neutrino production mechanism is active and efficient. Eq. (\ref{equ:diffmaster}) also includes the speed of light, $c$, the Hubble constant, $H_0$, and the fractions of cosmic energy density in dark matter and dark energy, $\Omega_M\simeq 0.3$, and $\Omega_\Lambda\simeq 0.7$. 

In computing the expression in \equ{diffmaster}, $z_{\mathrm{max}}=6$ is used, its value influences the result only weakly. We also choose $M_{\mathrm{min}}=2 \,  10^6 \, \msun$ and $M_{\mathrm{max}}=5~ 10^7\msun$, which is justified by the estimated masses in Table \ref{tab:observations}, when considering their uncertainty (at least a factor of two). The upper cutoff $M_{\mathrm{max}}$ is also expected because tidal disruption becomes increasingly inefficient for increasing $M$, see, \eg\ \citet{Kochanek:2016zzg}.    

The integration in $M$ is approximated by a discrete sum over three mass bins. 
Specifically, we work under the assumption that the entire \td\ population is represented, although roughly, by the three neutrino-detected \tds, each of which corresponding to a different value of $M$. For each model, our benchmark scenario assumes that the three \n\ spectra found for \dsg, \fdr\ and \aalc\ contribute equally to the diffuse flux, so they are assigned equal weights: $(w_1,w_2,w_3)=(1/3, 1/3, 1/3)$. This choice is the best inference that can be obtained from observations. It is also plausible theoretically: considering the three mass values in Table \ref{tab:observations} with uncertainties, it is possible that the observed \tds\ may fall in the mass bins $M/\msun=[2~ 10^6, 4~ 10^6], [4~ 10^6, 10^7], [10^7, 4~ 10^7]$, which correspond to equal \td\ rates for our chosen $\dot{\rho}(z,M)$. To describe the uncertainty on the \n\ spectrum and normalization, we also vary over all the possible weights $(w_1,w_2,w_3)$, thus obtaining an envelope of curves with the purpose to quantify the uncertainty of how representative each TDE is for the full population. 

\begin{figure}[t]
    \centering
    \includegraphics[width=0.4\textwidth]{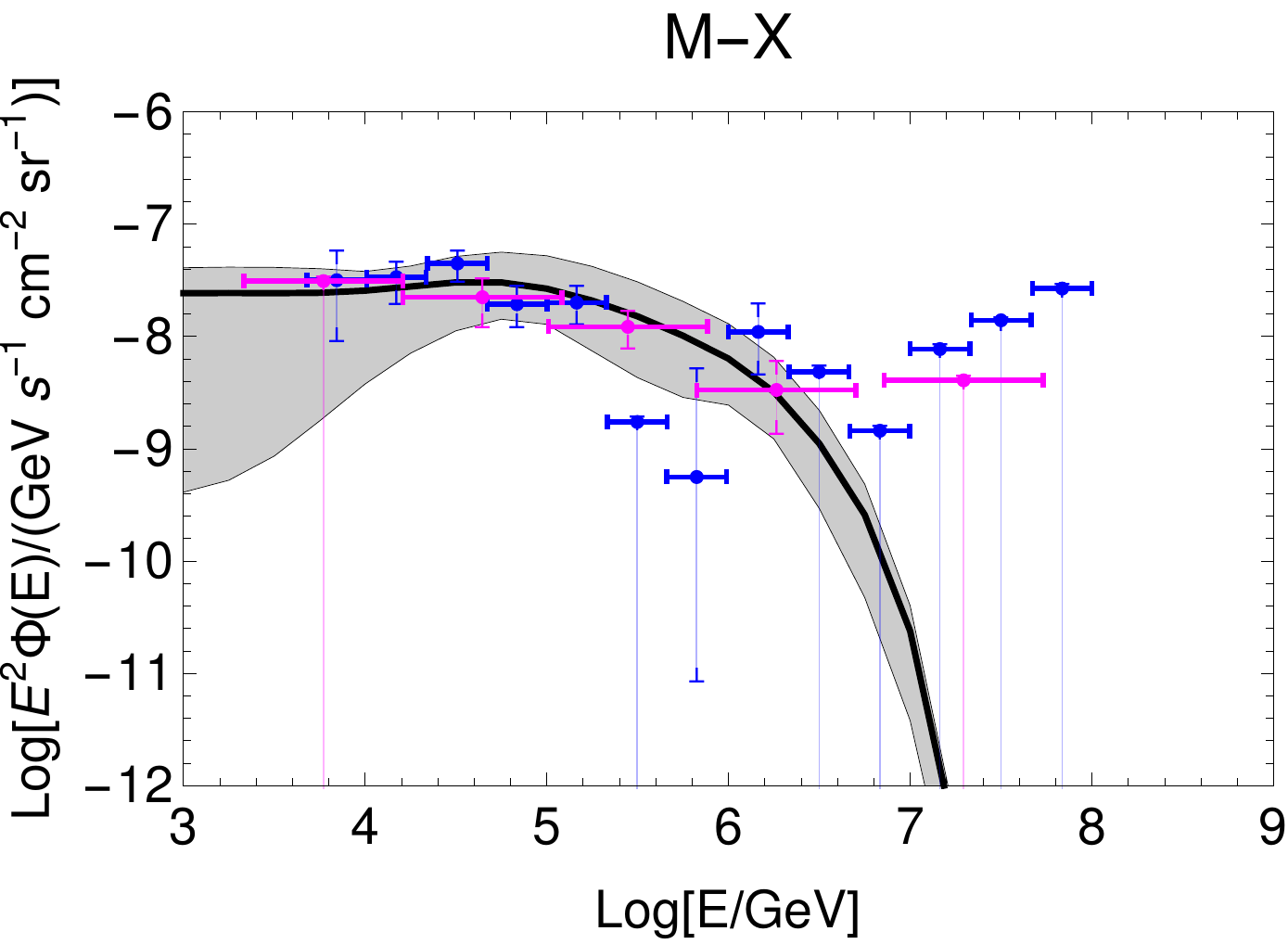}
    \includegraphics[width=0.4\textwidth]{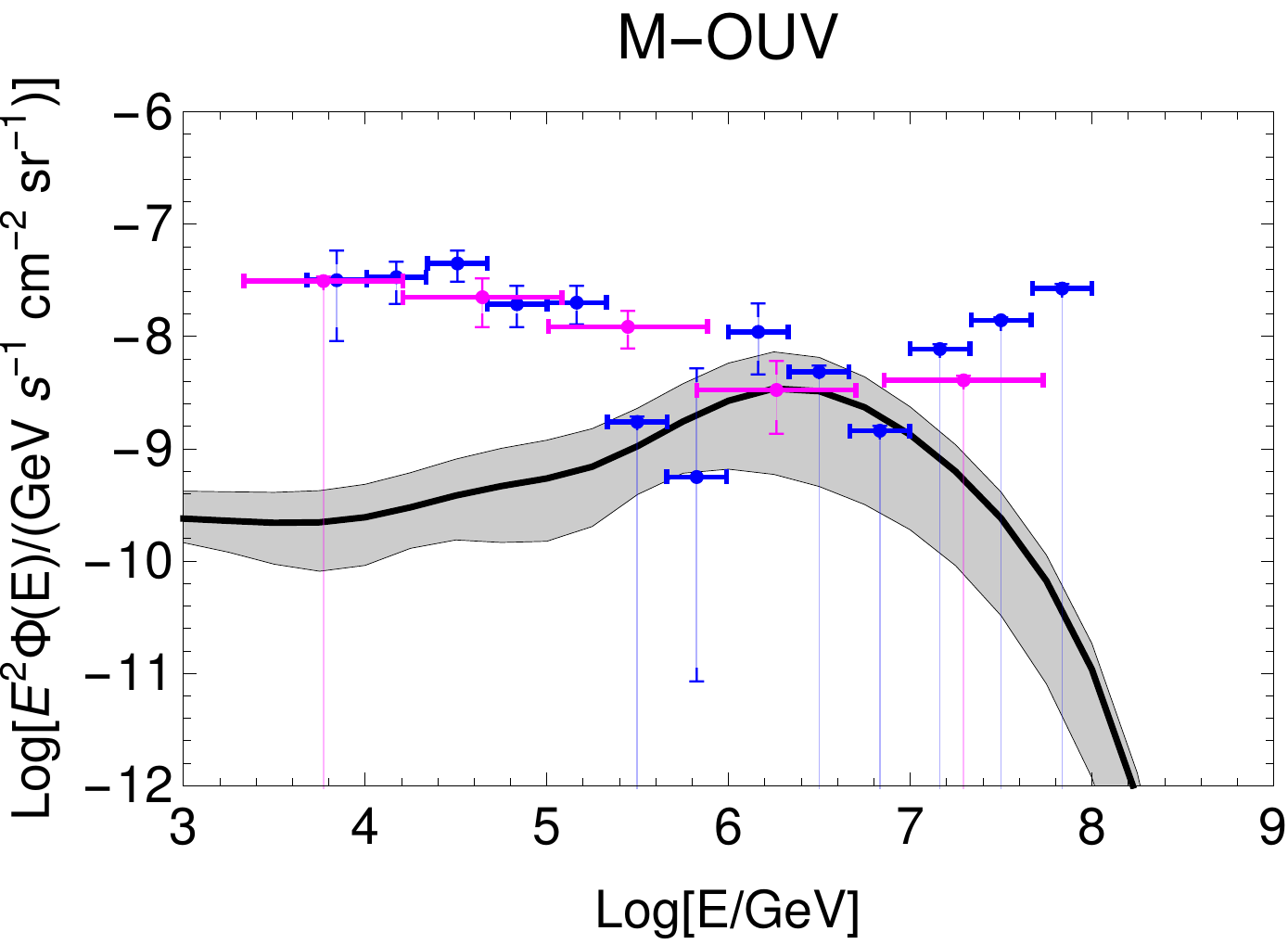}
    \includegraphics[width=0.4\textwidth]{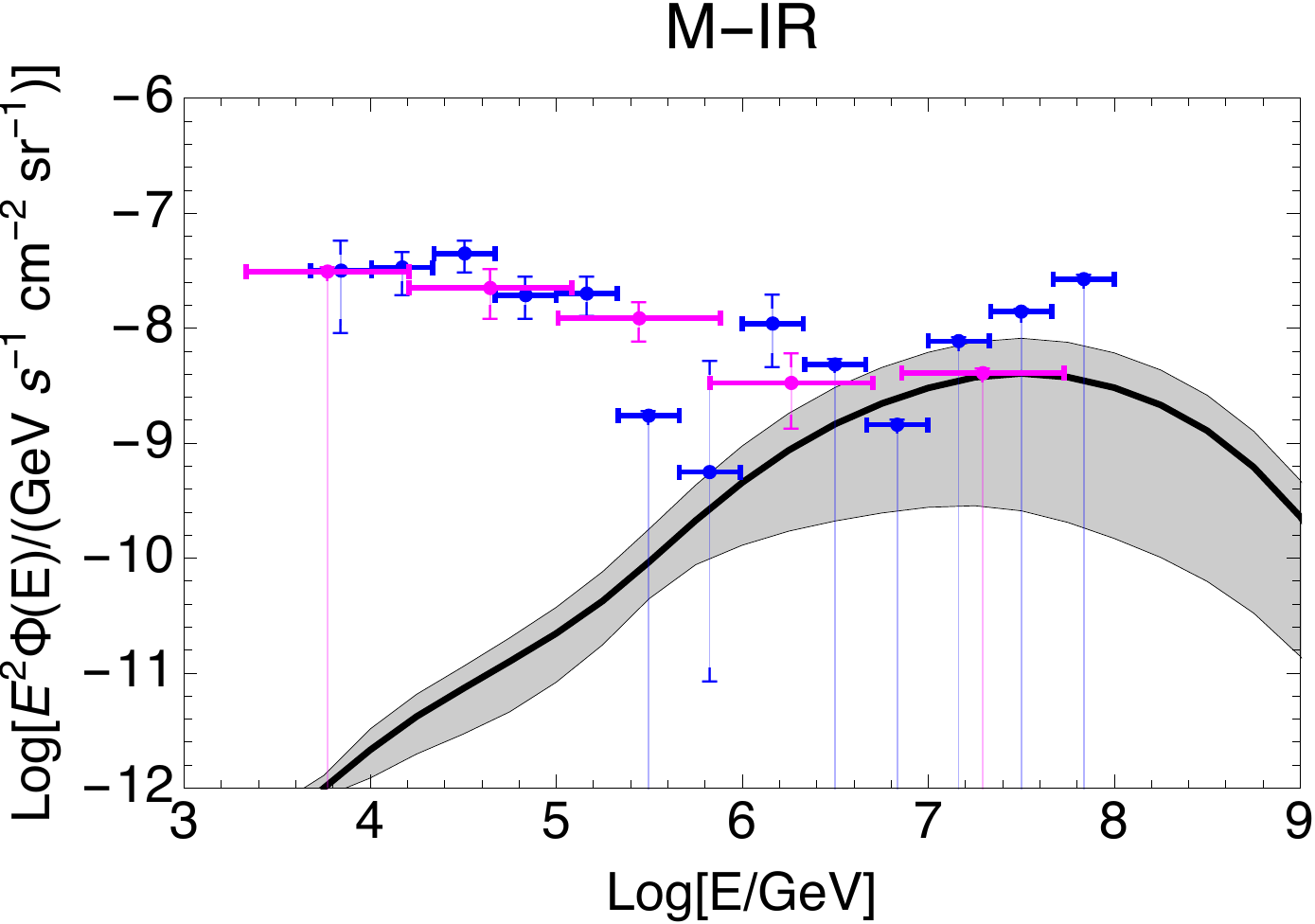}
    \caption{$\numu + \barnumu$ diffuse fluxes for the three models, for the case where the three single-\td\  spectra contribute with equal weights (solid curves) or arbitrary weights are allowed (shaded areas). The fractions of contributing TDEs have been set to the values given in \equ{etaval} to reproduce the observed diffuse flux. Data points represent the IceCube measurements from cascades \citep{IceCube:2020acn} (blue, data with narrower energy bins) and tracks \citep{IceCube:2021uhz} (magenta, data with larger energy bins).   
    }
    \label{fig:diffuseflux}
\end{figure}

The resulting $\numu + \barnumu$ flux is shown in \figu{diffuseflux} for the benchmark case (central curve) as well as the varying weights one (shaded area). Also shown are the corresponding flux measurements from cascade-like \citep{IceCube:2020acn} and track-like  \citep{IceCube:2021uhz} events at IceCube.   In the figure,
the fraction of neutrino-emitting \tds, $\eta \lesssim 1$, has been adjusted to reproduce the data;  chosen values are 
\begin{equation}
    \eta= \begin{cases}
  10^{-0.5}, \, 10^{-2.3}, \, 10^{-2.0}  & \text{ for~~} \varepsilon_{\mathrm{diss}}=0.2~ \\
10^{0.1}, \, 10^{-1.7}, \, 10^{-1.4} & \text{ for~~} \varepsilon_{\mathrm{diss}}=0.05~
\end{cases} \, ,
    \label{equ:etaval}
\end{equation}
where the three values listed in each case refer to M-X, M-OUV and M-IR respectively. Here we have used the degeneracy between $\eta$ and  the dissipation efficiency $\varepsilon_{\mathrm{diss}}$: we observe that  $ Q_\alpha(E(1+z),M)\propto \varepsilon_{\mathrm{diss}}$, and therefore $\Phi_\alpha \propto \eta \, \varepsilon_{\mathrm{diss}}$.
This implies that the data can be reproduced with a more moderate requirements for the dissipation efficiency into non-thermal protons over the whole TDE population, at the expense of a larger fraction of neutrino emitting TDEs (which could be as large as one). There is also a degeneracy with $M_{\mathrm{min}}$: due to the negative evolution of the \td\ rate with mass,  $\Phi_{\alpha}$ increases when lowering $M_{\mathrm{min}}$. For $M_{\mathrm{min}}\sim 10^5\msun$ an order of magnitude increase is expected with respect to our results (see, \eg, \citet{Lunardini:2016xwi}), thus leading to lower requirements for $\eta \, \varepsilon_{\mathrm{diss}}$.  
From  \figu{diffuseflux} and the comparison with the measured spectrum, we conclude that, for M-OUV and M-IR, \tds\ cannot be the main contributors to the astrophysical flux observed by IceCube, but they may significantly contribute at the highest energies $E \gtrsim 1$ PeV.  In contrast, for M-X the predicted spectrum reproduces the data over three orders of magnitude of energy.
We note that this conclusion depends strongly on the spectrum weights $w_i$; in the case where the spectrum for \dsg\ (which is more suppressed at low energy compared to the others, see \figu{mx}) carries a large weight, the observed flux at $E \lesssim 0.1$ PeV can not be accounted for by \tds. 

It is an interesting question if the chosen values of $\eta$ are roughly consistent with the number of neutrino-TDE associations that have been found through neutrino follow-up programs.  If we consider the rate of these associations to be about one per year, generically one expects  
$\eta \simeq 1 \, \mathrm{yr}^{-1}/(\dot N \, N_{\mathrm{GFU}})$, 
where $\dot N$ is yearly rate of TDEs that can -- at least in principle -- be found by standalone or follow-up observations (``observable'' \tds ), $N_{\mathrm{GFU}}$ is the predicted neutrino event rate from our models, and we assume (optimistically) that most neutrino events are followed up. 
An estimate of $\dot N$ can be obtained by integrating $\dot \rho(z,M)$ in the quoted mass range, and up to a redshift that roughly matches the reach of current instruments. 
We obtain $\dot N \simeq 10^4 \, \mathrm{yr}^{-1}$ for $z<0.3$, and $\dot N \simeq 400 \, \mathrm{yr}^{-1}$ for $z<0.1$, which serve as the range of uncertainty for this estimate from the instrument redshift threshold only\footnote{Our results for $\dot N$ imply that the number of {\em observable} TDEs per year is much larger than the number of actually {\em observed} TDEs  (perhaps a few tens) per year. The difference between the two rates is explained by the duty cycle and field of view of the instruments, and the difficulty to classify events as TDEs -- even the nearby ones -- because of the instrument threshold.}.
Taking the average values for $N_{\mathrm{GFU}}$ over the three TDEs (see \figu{cartoon_MX}, \figu{cartoon_MOUV}, \figu{cartoon_MIR}, right tables), one would expect the following ranges for $\eta$: for $\varepsilon_{\mathrm{diss}}=0.2$, the intervals $\eta \in [10^{-2.6},10^{-1.2}]$, $[10^{-3.7},10^{-2.3}]$, and $[10^{-2.4},10^{-1.0}]$
for M-X, M-OUV, and M-IR, respectively, and for $\varepsilon_{\mathrm{diss}}=0.05$ the intervals $\eta \in [10^{-2.0},10^{-0.6}]$, $[10^{-3.1},10^{-1.7}]$, and $[10^{-1.8},10^{-0.4}]$. 
Comparing these expectations to reproduce the number of TDE associations with the numbers to reproduce data in the respective energy ranges in \equ{etaval}, we can estimate the expected contribution to the diffuse flux from the ratio of these numbers: $0.8\%-20\%$, $\gtrsim 4\%$, and $\gtrsim 38\%$ for M-X, M-OUV, and M-IR, respectively, where the dependence on $\varepsilon_{\mathrm{diss}}$ cancels. This means that the
diffuse neutrino flux for M-X shown in \figu{diffuseflux} would probably lead to too many neutrino neutrino-TDE associations, while the other two are plausible.
 For comparison, a range between 5\% and 59\% of the diffuse flux is given in~\citet{Bartos:2021tok} at the 90\% CL; this range is roughly consistent with our estimates, given the systematic uncertainties. Note that the number of astronomically (in the electromagnetic bands) observable neutrino-emitting TDEs is given by $\eta \cdot \dot N$ in this approach, which ranges between 2 and 40 per year (8 and 160 per year) for $\dot N \simeq 400 \, \mathrm{yr}^{-1}$ and $\varepsilon_{\mathrm{diss}}=0.2$ (0.05). That ranges are roughly consistent with the number of interesting TDE candidates found in \citet{vanVelzen:2021zsm} selected by the strength of the dust echoes, which supports  the hypotheses that the neutrino-emitting TDEs and the TDEs with strong dust echoes could indeed be the same populations. It is also interesting to note that, after the discovery of the jetted TDE AT2022cmc, the recently derived fraction of TDEs having jets in the percent range \citep{Nature:AT2022cmc} is consistent with our estimated ranges for $\eta$, which means that also the neutrino emitting TDEs and jetted TDEs could be the same populations; see \App~\ref{app:jetted} for details.

\section{Comparison and discussion}
\label{sec:discussion}

\begin{table}[t]
    \centering
    \begin{tabular}{|l|ccc|}
    \hline
     {\bf Model criterion} & {\bf M-X} & {\bf M-OUV} & {\bf M-IR} \\
    \hline
    Accelerator: Scale comparison & $R \simeq R_{\mathrm{acc}} \gtrsim R_{\mathrm{BB}}$ & $R \simeq R_{\mathrm{BB}} \gtrsim R_{\mathrm{acc}}$ &  $R \simeq R_{\mathrm{dust}} \gtrsim R_{\mathrm{acc}}$ \\
     \hspace*{0.5cm}Required $\varepsilon_{\mathrm{diss}}$ & $\gtrsim 0.1-0.2$ & $\gtrsim 0.03$ & $\gtrsim 0.1$  \\
    \hspace*{0.5cm}Wind/outflow models & Challenged by $\varepsilon_{\mathrm{diss}}$, $t_{\mathrm{ad}}$ & Unlikely ($R_{\mathrm{acc}}>R$, $\eta_{\mathrm{acc}}$) & Challenged by  $\varepsilon_{\mathrm{diss}}$ \\
    \hspace*{0.5cm}Off-axis jet & Yes & No (no isotropization in $R$) & Yes \\
    \hspace*{0.5cm}Core models (disk, corona) & Not described
    & Yes, but $E_{p,\mathrm{max}}$? & Not fully described \\
 \hline
    Main targets & X-rays, protons & Optical-UV blackbody & IR from dust echo \\
    Observational evidence/correlation & X-ray signals & High $L_{\mathrm{BB}}$ & Dust echoes \\
    Origin of neutrino time delay & Diffusion (high $B$) & Unrelated to size of system & Dust echo travel times \\ 
    \hline
    Description neutrino time delay & Intermediate & Poor & {\bf Good} \\
    Neutrino event rate & Low & {\bf Intermediate-High} & Low \\
    Required $E_{p,\mathrm{max}}$ & {\bf Moderate} & Intermediate & Ultra-high \\
    Neutrino energy & {\bf Matches} & Somewhat high & Very high \\
    Neutrino spectral time evolution &  {\bf Matches} & Right direction & Wrong direction \\
    \hline
    Diffuse flux spectral shape & {\bf Matches} & High $E$ only & Highest $E$ only \\
    Diffuse flux contribution &  $0.8\%-20\%$ &  $\gtrsim 4\%$ (high $E$ only) & $\gtrsim 38\%$ (highest $E$ only) \\
    \hline
    \end{tabular}
    \caption{Qualitative comparison of different models regarding different aspects; best matches are boldface. }
    \label{tab:comptable}
\end{table}

Let us take a comparative look at the three models we have proposed; 
a first question that arises is which model, if any, is favored by observations. 
We give a qualitative comparison of the three models in \Tab~\ref{tab:comptable}. From the table, it appears that at present there is no clear preference for a single model. In terms of neutrino signal, M-X describes the neutrino energy very well and the required $E_{p,\mathrm{max}}$ is moderate; M-OUV can most easily describe the expected neutrino fluence, and therefore has the best requirement for the transfer efficiency of accretion power into non-thermal protons; M-IR describes the neutrino time delay best. Major drawbacks are the poor time delay description in model M-OUV, and the ultra-high proton and neutrino energies in model M-IR, so that, overall M-X seems to be a very attractive choice. However, M-X can only describe up to 20\% of the diffuse neutrino flux as estimated from the neutrino-TDE associations, and the neutrino signal from \aalc\ may not be driven by X-rays due to the low observed X-ray luminosity. The best candidate for the acceleration region might be an off-axis jet for model M-X, an accelerator in the core for M-OUV (if  high enough proton energies can be reached), and an outflow, wind or jet for M-IR. 

\begin{figure}[t]
\begin{center}
\begin{tabular}{ccc}
\includegraphics[width=0.3\textwidth]{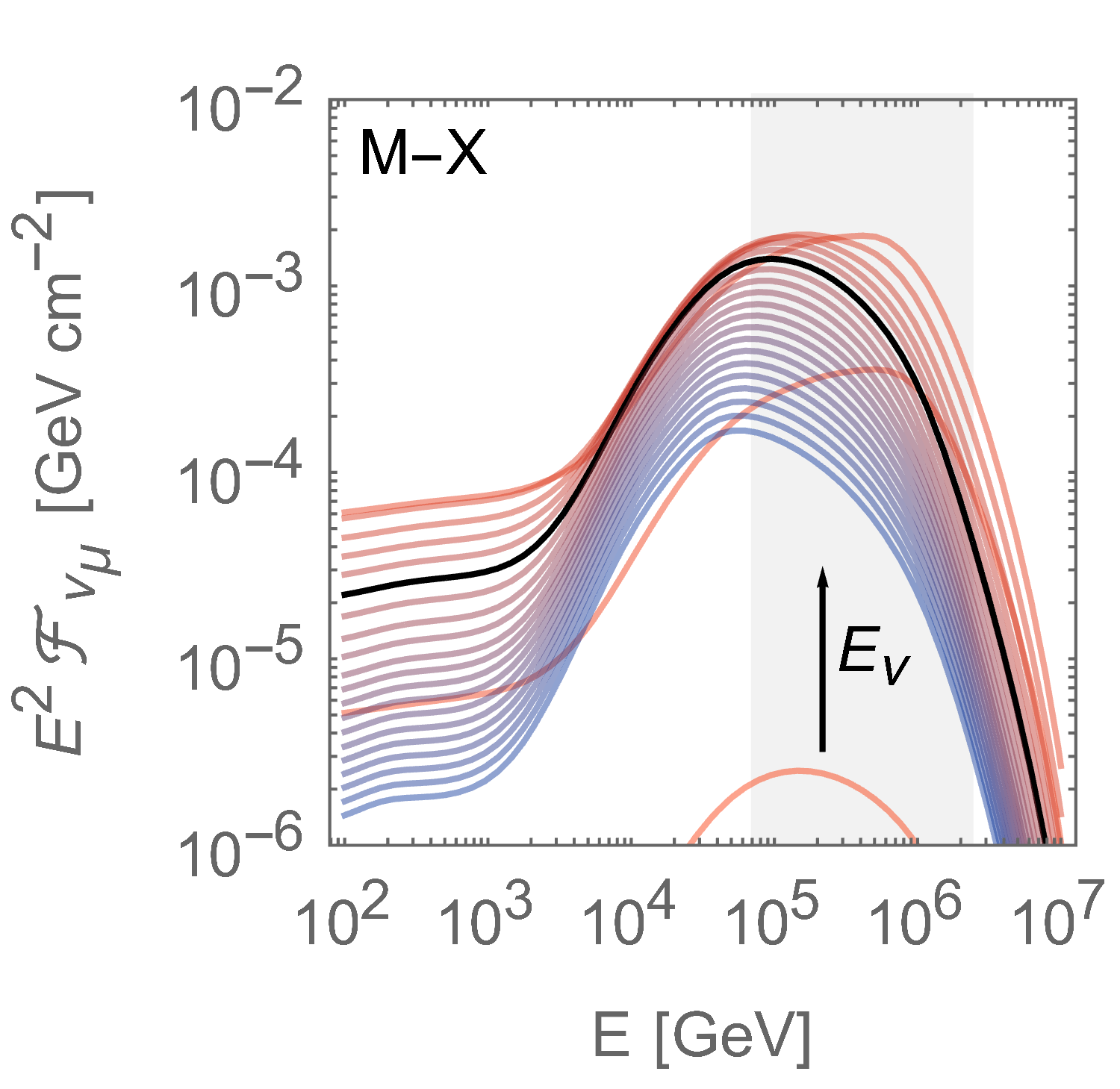} &
\includegraphics[width=0.3\textwidth]{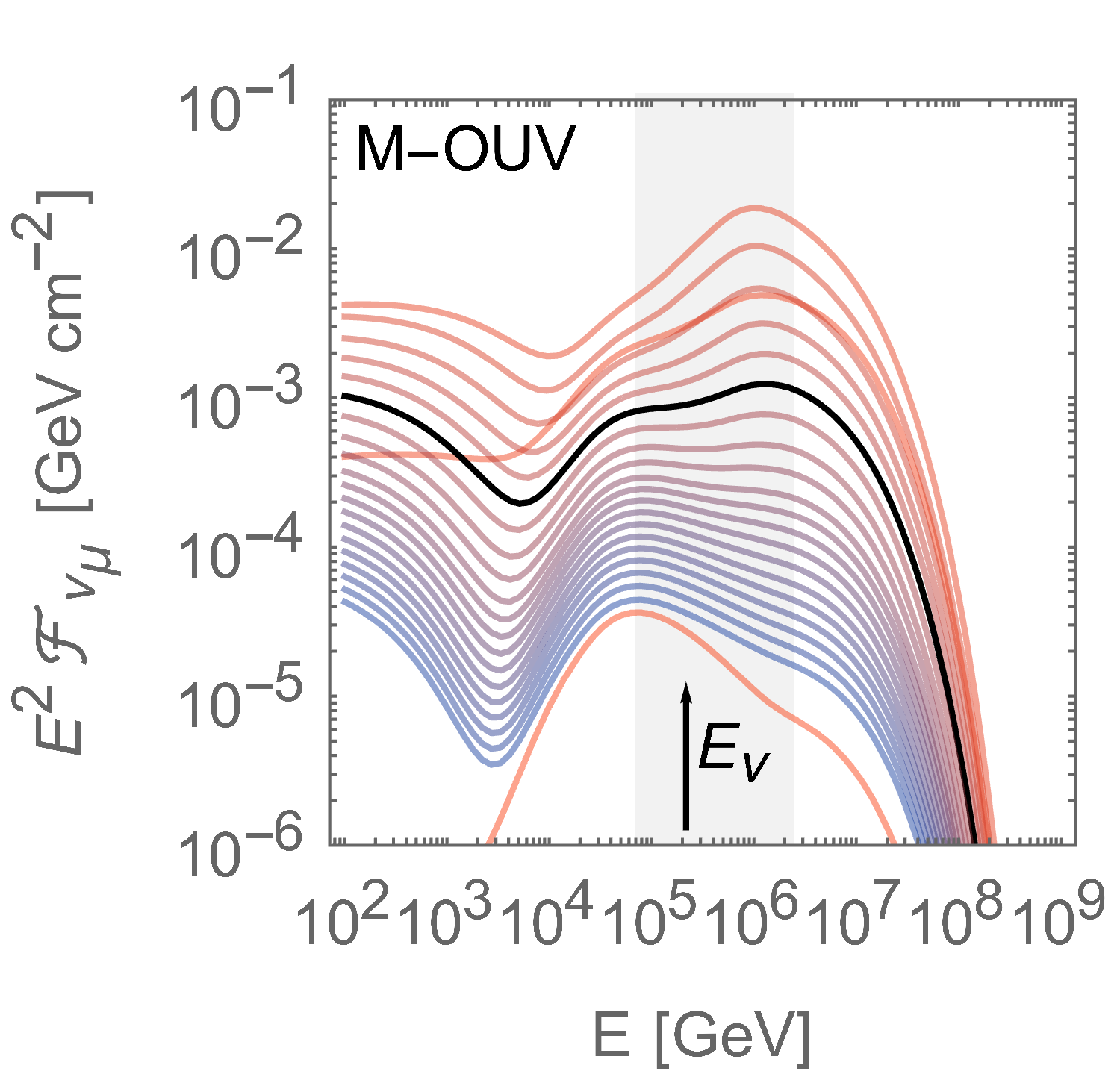} &
\includegraphics[width=0.3\textwidth]{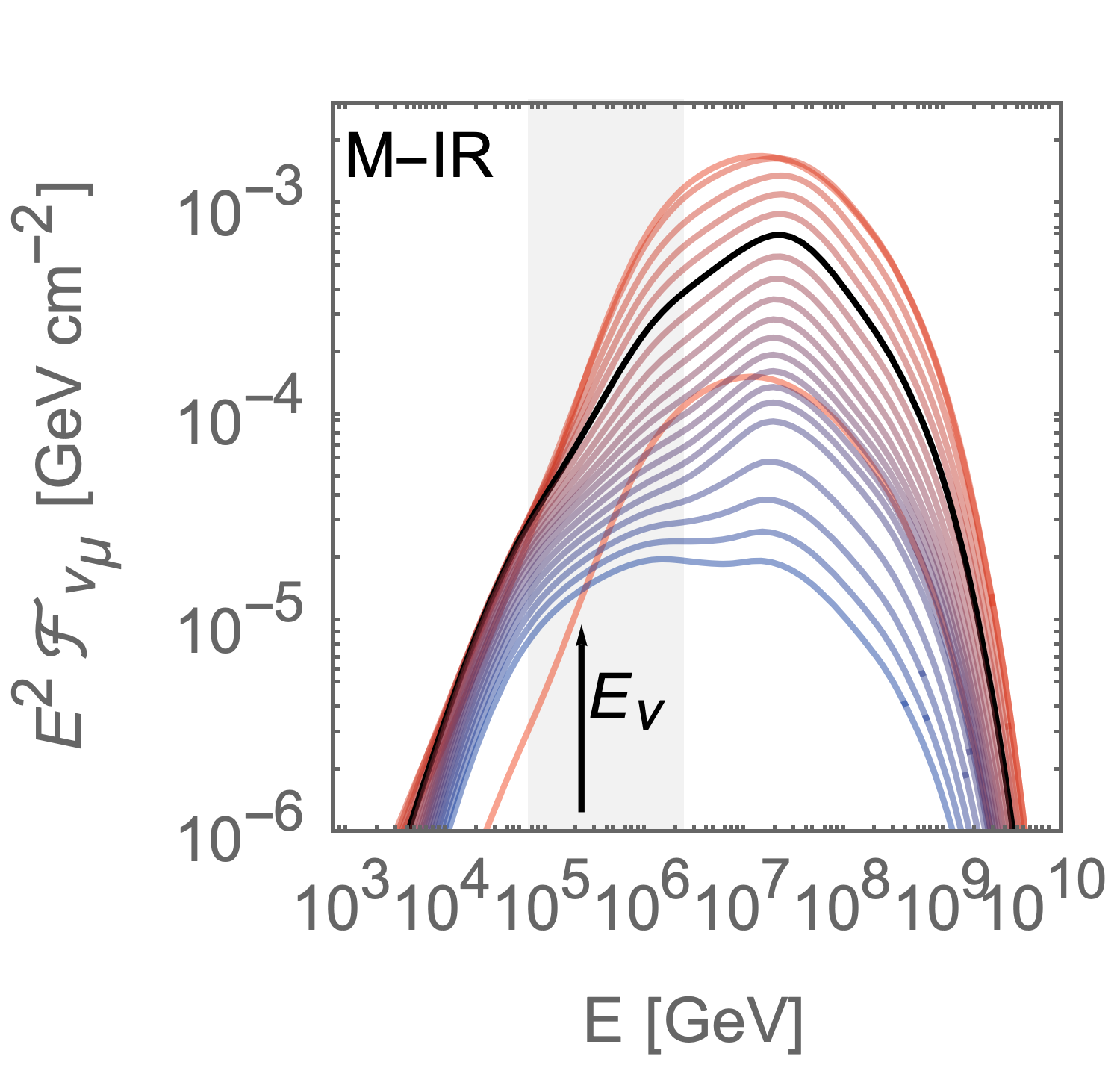} \\[0.3cm]
\multicolumn{3}{c}{\includegraphics[width=0.9\textwidth]{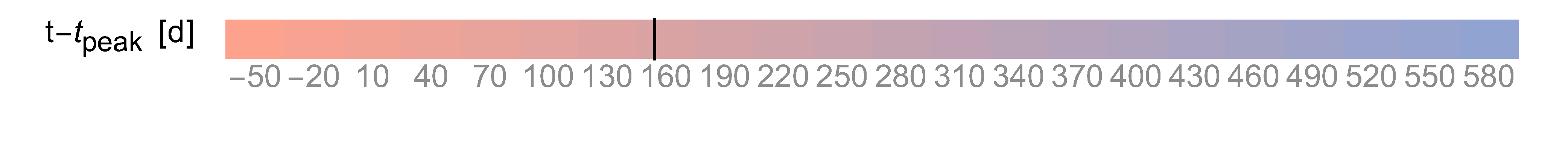}} 
\end{tabular}
\end{center}
\caption{\label{fig:timeevol} Time evolution of the emitted neutrino fluence  as a function of the observed energies for the three different models for \dsg\, compared. Note that the individual shown fluences for the time slices add up to the total fluence. Black curves are closest to the actual neutrino emission, likely neutrino energies are marked as described in the caption of \figu{mx}.}
\end{figure}

As was noted before, all three neutrino-TDE associations require SMBH masses around $10^7 \, M_\odot$, which is high compared to the mean of the observed \td\ population. Our models are roughly consistent with this scaling with $M$. Indeed, for M-OUV, both the proton injection rate $L_p$ (by construction) and the BB luminosity (from observations/theoretical expectations, see Table~\ref{tab:observations}) are proportional to $ L_{\mathrm{Edd}} \propto M$, which means that one expects $L_\nu \propto L_p \times \tau_{p \gamma} \propto L_{p} \times L_{\mathrm{BB}}  \times T^{-1}\propto L_{p} \times L^{3/4}_{\mathrm{BB}}  \propto M^{7/4}$ (where eq. (\ref{equ:taumx}) and the Stefan-Bolzmann law were used) in the $p\gamma$ optically thin case, saturating to  $L_\nu \propto L_{p}  \propto M$ in the optically thick case (where all proton energy is effectively transferred into neutrinos) --  which means that the neutrino-TDE associations will be likely dominated by the upper end of the SMBH range even if the differential TDE rate scales as $M^{-1.6}$, see discussion in~\citet{Lunardini:2016xwi}. 
For M-IR, the same argument holds, since the intensity of the dust echo is proportional to the one in OUV. Instead, predicting a scaling with $M$ for M-X is more difficult.
For \aalc, the \n\ production is dominated by $pp$ interactions,  for which the same scaling of the other models applies. For the other two \tds, one could apply the prediction by  \citet{Mummery:2021nqy} (see also \citet{Mummery:2019ggj}) that $L_X$ should be nearly constant  over a wide mass range, leading to $L_\nu \propto M$ even in the optically thin case and to the same qualitative agreement with the three measured \bh\ masses (note however, that  the X-ray spectrum would have a non-trivial dependence on $M$ \citep{Mummery:2021nqy}, which would affect the \n\ spectrum). 

To elaborate on the consistency between the models and detected \n\ events, 
let us examine the time-dependent evolution of the neutrino spectrum, which could be complicated. Especially if different target photon spectra contribute to the neutrino production, we have seen that the interaction rates will be very different for these; this means that \eg\ neutrinos from OUV interactions will be closely following the proton injection because the system is optically thick, whereas neutrinos from X-ray and IR interactions are intrinsically produced later due to the calorimetric behavior. Therefore, there is a connection between the expected neutrino energy and the time delay. We illustrate this in \figu{timeevol} for the three different models for the example of \dsg , where the time evolution of the neutrino spectrum is shown  (from red--early to blue--late, black curves refer to the time of the neutrino emission). For M-X, the early spectrum from OUV interactions peaks at higher energies than the spectrum at the actual time of the neutrino emission; in all cases the spectrum matches well. For M-OUV, the overall neutrino emission will be dominated by earlier times and relatively high neutrino energies within the possible range, whereas at the time of neutrino emission the spectrum is relatively flat in the expected (gray-shaded) range; the spectral evolution thus is in the right direction. For M-IR, it is actually the early emission (from the OUV target) which peaks closer to the observed neutrino energy, whereas at the time of the neutrino emission, the spectrum peaks at too high energies; so, the temporal evolution of the spectrum actually increases the tension with observations. These qualitative observations are also listed in our \Tab~\ref{tab:comptable}.

At a more detailed level, one could wonder which of the three observed \tds\ is best described by our models. 
We find that for \aalc\ the neutrino fluence and time delay are well reproduced because of its relatively large SMBH mass (resulting in higher luminosities), low redshift, and slower decline of the BB luminosity. \dsg\, on the other hand, has a faster declining BB luminosity, which makes the description of the neutrino time delay more difficult because non-thermal proton injection will peak early. While \fdr\ exhibts high neutrino luminosities in the SMBH frame, the neutrino fluence at Earth is suppressed due to the high redshift. Interestingly, for \aalc\  very high point source event rates are found for models M-X and M-OUV, which leads to the conclusion that (subject to Poissonian fluctuations) the source could have even been independently found as neutrino point source for large enough $\varepsilon_{\mathrm{diss}}$; instead, for the other two \tds\ the Eddington bias argument has to be invoked.

High neutrino fluences, however, imply that gamma-rays from accompanying neutral pion decays will be abundantly produced, which means that gamma-ray observations may constrain the predicted neutrino fluxes. The electromagnetic cascade triggered by gamma-gamma pair production depends on the time-evolution of the photon density across the spectrum, potentially including photon components not included in our models. In the most conservative optically thick (to $\tau_{\gamma \gamma}$) case, the cascaded gamma-ray flux is expected to be at the level of the neutrino flux typically peaking below the pair production threshold, which is about 5 GeV if X-rays are abundant enough, and 500~GeV otherwise (such as for \aalc ). If one compares the neutrino fluxes to the gamma-ray limits in \cite{vanVelzen:2021zsm}, one finds rough consistency for most TDEs and models, but the expected gamma-ray fluxes are potentially observable in some cases, especially for M-OUV (all TDEs) and \dsg\ (all models). While detailed models are beyond the scope of this study, one may expect that $\varepsilon_{\mathrm{diss}}$ could be constrained.

Finally, let us comment on how our proposed mechanisms could apply to other classes of sources. A natural comparison is with AGN. In \citet{vanVelzen:2021zsm} it is concluded that AGN must be less ``efficient particle accelerators'' than TDEs because, in the electromagnetic channel, steady AGN emission outshines that of  TDEs at least by two orders of magnitude -- and the same is not observed in neutrinos, despite the similarities between AGN and TDEs (for example, all AGN have large regions with hot dust which could produce targets for neutrino production). We  anticipate that the comparison is in fact more complex than an electromagnetic output comparison. 
First of all, the acceleration of protons to such high energies may only take place in components present in one source class, such as the debris stream-return stream collisions in TDEs, which means that the accelerator may simply not be present in AGN (or have very different properties). Second, since the neutrino luminosity (for p$\gamma$ interactions in the optically thin case) scales as $L_\nu \propto L_p \times \tau_{p \gamma}$, where  $\tau_{p \gamma}$ strongly depends on the size of the radiation zone,  efficient proton acceleration (powering $L_p$) is only one part of the problem, the other is how efficient these non-thermal protons (if they exist) can transfer energy into neutrinos (magnitude of $\tau_{p \gamma}$).  To illustrate that, let us consider jets, which could be realized in both source classes. It is known that AGN blazars (jets pointing towards us) suffer from low $\tau_{p \gamma}$ for parameters required to describe the spectral energy distribution (see \eg\ \citet{Gao:2018mnu} for TXS 0506+056, and \citet{Oikonomou:2022gtz} for a more general discussion). This needs to be compensated by a correspondingly larger $L_p$, i.e., large baryonic loading often related to super-Eddington accretion rates, which are sometimes perceived to be unrealistic. Our models are qualitatively different: $\tau_{p \gamma}^{\mathrm{cal}}$ in the relevant radiation zone is comparatively large (which means that TDEs are indeed efficient neutrino sources), and hence the required baryonic loading (if defined as energy injected into protons versus BB) of about 30-100 (even for the large $\varepsilon_{\mathrm{diss}}=0.2$) is more comparable to expectations for GRBs as UHECR sources than AGN, see \eg~\citet{Heinze:2020zqb}. This means that TDEs may  (i) host more efficient proton-accelerating sites, (ii) have more compact radiation zones, or (iii) support a more effective proton confinement than AGN. That, on the other hand, implies that models clearly inspired by AGN physics, such as the corona core model~\citep{Murase:2020lnu}, are challenged by this argument -- and must require very different parameters in AGN and TDE to explain that observed difference.

\section{Summary and conclusions}
\label{sec:summary}

We have studied fully time-dependent quasi-isotropic neutrino production models for the three TDEs associated with high-energy astrophysical neutrinos, postulating that the observed neutrino time delays with respect to the BB peak come from the physical size of the post-disruption system, such as confinement of protons in magnetic fields over a large enough region, or propagation time delays. We have pointed out that the dominant photon target for proton interactions depends on the available maximal proton energy provided by the acceleration region; we have not specified the accelerator explicitly, but instead have parameterized it by the maximal proton energy and proton injection luminosity of a non-thermal spectrum (with power law index two); examples could be off-axis jets, hidden winds, shocks from outflow-environment interactions or stream-stream collisions. For one of the  models (M-OUV) disk or corona could be acceleration sites, too, if high enough proton energies can be reached.

We have focused on observations common to the three TDEs, which, apart from the neutrino time delays, are: a) X-ray signals, b) relatively high SMBH masses, and, correspondingly, also relatively high BB luminosities, and c) strong dust echoes. Our models consequently have adopted X-ray (model M-X), optical-UV BB (model M-OUV), and infrared  (model M-IR) photons as main interaction targets, selected by the maximally available proton energy, targeting the question of what the smoking gun signature for the neutrino production actually is. A qualitative comparison is given in \Tab~\ref{tab:comptable}. Model M-X describes the observed neutrino energies and time delays well due to the confinement of moderate-energy protons if the unobscured X-ray luminosity is roughly constant as a function of time; model M-OUV describes the highest neutrino event rates at the expense of small neutrino time delays; model M-IR provides a good description of the neutrino time delays because of the correlation with dust echoes with similar delays and it may actually power the diffuse neutrino flux at the highest energies, at the expense of very high proton and neutrino energies. Note that TDEs may also be the sources of UHECRs if model M-IR can be established, but a more quantitative approach requires further study.

Since our models predicts a neutrino luminosity roughly scaling as $L_\nu \propto M^{(1-2)}$, it not surprising that the observed neutrino-TDE associations involve \bh\ that are more massive than the mean of the observed \td\ population. From that perspective, the newly found TDE \aalc , which was associated with the highest SMBH mass $M$ (subject to large uncertainties) and the lowest redshift, is expected to produce the highest neutrino fluence, which we have seen in all models. In fact, in two models the event number using the point source effective area was larger than one if the dissipation efficiency $\varepsilon_{\mathrm{diss}}$ is high, which means that this source could be seen in (transient) multiplet or point source searches, whereas the detection of the other two was predicted to be a matter of chance (from a larger sample of sources with low predicted event rates each, invoking the Eddington bias). Note, however, that the information on \aalc\ is sparse, which means that there are larger uncertainties (than for the other two TDEs). Especially in this case, the search for/comparison to additional electromagnetic signatures, \eg\ from secondaries produced in the photo-pion production,  which may also follow the time-dependence of the neutrino spectra, could be interesting.

A major limitation of our approach are relatively large required dissipation efficiencies $\varepsilon_{\mathrm{diss}} \gg 0.05$ from mass accretion rate into non-thermal protons for the individual  TDE models M-X and M-IR, whereas M-OUV can tolerate smaller efficiencies due to the high BB target photon density. For example, the requirement of high $\varepsilon_{\mathrm{diss}} $ poses a challenge to  outflow models if $v\lesssim 0.5 \, c$, and to off-axis jetted models if the dissipation efficiency of kinetic energy into non-thermal particles (e.g. in internal shocks) is less than about 25\%. Our diffuse flux computation has demonstrated that the average $\varepsilon_{\mathrm{diss}}$ of the whole population could be lower, at the expense of a larger fraction of neutrino-emitting TDEs.  As far as the origin of the neutrino time delays is concerned, its description has been especially challenging by our assumption that the non-thermal proton injection follows the mass accretion rate;
we have proposed a calorimetric approach paired with a relatively low (free-streaming) optical thickness (models M-X and M-IR), or paired with time-delayed target photons (dust echo model M-IR). Note that the magnetic confinement of protons has the advantage that protons emitted in different directions isotropize, such as from off-jets. Other possible reasons for the neutrino time delays include a time evolution of the proton injection different from the mass accretion rate, or a transition of the disk state in core models. 


We conclude that a decision of the neutrino production model based on the available information cannot be made; future observations will show if X-ray signals are associated with neutrino from TDEs (points towards M-X), or if dust echoes are observed (pointing towards M-IR); in both these cases time delays of the neutrinos are expected as well. If, on the other hand, BB-luminous TDEs with neutrinos close to the BB peak are found, M-OUV will be preferred. Note that other signatures initially gauged interesting for \dsg\ \citep{Stein:2020xhk}, such as the outflow and radio signals, may actually be of secondary importance for the neutrino production in the light of \fdr\ and \aalc . Important clues will also come from the observed neutrino energies, which will have to be scrutinized with more realistic spectra; minor correlations between neutrino arrival time and neutrino energy are also expected in the models. Depending on the scenario, protons, such as in an outflow or the debris stream, may also be a target for neutrino production; in our cases, these have not affected the qualitative conclusions, but may help to describe the soft diffuse neutrino flux at the low energies. 
A more challenging questions may be the origin of the accelerated protons: can these be associated with other non-thermal signature in the electromagnetic spectrum which will allow for an identification of the acceleration region, or will the origin of the non-thermal protons remain a mystery? 

\subsubsection*{Acknowledgments}

We would like to thank Claire Guépin, Marek Kowalski, Simeon Reusch, Xavier Rodrigues, Annika Rudolph, and Sjoert van Velzen for useful comments and discussions.
CL acknowledges funding from the National Science Foundation grant number PHY-2012195. 

\appendix

\section{Discussion of jetted models (on-axis, off-axis)}
\label{app:jetted}

 The isotropic models presented in this study are very different from jetted models (with jets pointed towards us, or slightly off-axis), see  \citet{Winter:2020ptf,Liu:2020isi} (\dsg)  and \citet{Reusch:2021ztx} (\fdr). 
The major limitation of the quasi-isotropic emission models is the very large fraction of energy has to go into non-thermal protons compared to models with a collimated emission, see~\citet{Winter:2021lyo} for a discussion. In jetted models $\varepsilon_{\mathrm{Comp}} \simeq 20\%$ of the total accretion power is assumed to go into the jet \citep{Dai:2018jbr}, and the transfer efficiency from kinetic energy to non-thermal radiation $\varepsilon_{\mathrm{NT}}$ can, depending on the outflow model, be around 10\% to 40\% (see e.g. \citet{Rudolph:2019ccl,Heinze:2020zqb}). Using $\varepsilon_{\mathrm{NT}} \simeq 25\%$  results in an overall  fraction $\varepsilon_{\mathrm{diss}} \simeq 0.05$ into non-thermal protons  -- which is at the lower end of the range proposed here; high isotropic-equivalent proton luminosities are then reached by relativistic beaming. Note, however, that we use the most conservative $E_{p,\mathrm{min}} \simeq 1 \, \mathrm{GeV}$ in all cases, which means that a large fraction of non-thermal protons cannot interact by $p \gamma$ interactions because they are below threshold; increasing $E_{p,\mathrm{min}}$ or harder acceleration spectra  would reduce the efficiency challenge for the individually observed TDEs. 
A big question for jetted models has been the non-observation of non-thermal internal radiation, and the interpretation of the radio observations for \dsg , which indicate that the jet may have been unusually narrow ($\theta \ll 1^\circ$)~\citep{Cendes:2021bvp}. This immediately raises the question why the neutrino-TDE associations can be so abundant, as most jetted TDEs would not be seen in our direction for such a narrow jet. 

Strongly off-axis jets acting as proton accelerators would be a possible solution to the puzzle, as the particle acceleration itself is known to work efficiently in astrophysical jets. In this case  escaping non-thermal protons need to escape on the scale within our radiation zone, and to be trapped within the calorimeter and isotropize -- which is in principle possible in our calorimetric models M-X and M-IR (see \figu{timescales_MX} and \figu{timescales_MIR}: energy ranges where $t^{-1}_{p,\mathrm{esc}}<t^{-1}_{\mathrm{fs}}$).
Similar ideas have been proposed on different scales, see \eg\ discussions in \citet{Fang:2017zjf,Rodrigues:2020pli}. Note that here the cosmic-ray escape mechanism from the jet is also critical: for example, neutron escape does not contribute as neutrons cannot be magnetically confined, and advective escape may not work at the relatively small scales $R$ proposed for M-X, whereas direct or diffusive escape of high-energy protons could work \citep{Baerwald:2013pu}, i.e., of protons for which the Larmor radius reaches the size of the accelerator.
The advantage of an off-axis jet is that the jet can serve as proton accelerator, while other radiation signatures of the jet are not expected or too weak if the viewing angle is large enough; in addition, all TDEs with jets would serve as neutrino sources (not only the ones pointing towards us), whereas the chance coincidence for a jet to point into our direction is at the percent level~\citep{Nature:AT2022cmc}. Since the required non-thermal proton luminosity is comparable to the typically expected physical (kinetic) jet luminosity, the jets would have to be very efficient to transfer kinetic energy into non-thermal particles for the three TDEs discussed here (and the jet collimation cannot be used to increase the isotropic-equivalent energy for an off-axis jet). For the diffuse flux, note that not all TDEs are expected to produce relativistic jets; fractions  of jetted  \tds\ of $\eta \simeq 0.1$, as assumed in \citet{Lunardini:2016xwi}, or, more conservatively, $\eta \simeq 0.01$, as inferred in \citet{Nature:AT2022cmc}, are compatible with the $\eta$-ranges derived in \Sec~\ref{sec:diffuse} from the number of neutrino-TDE associations for M-X and M-IR (for $\varepsilon_{\mathrm{diss}} \simeq 0.05$), which means that the neutrino-emitting \tds\ and jetted TDEs could be the same population.

\section{On the effect of adiabatic cooling}
\label{app:adiabatic}

 \begin{figure}[t]
\begin{center}
\includegraphics[width=\textwidth]{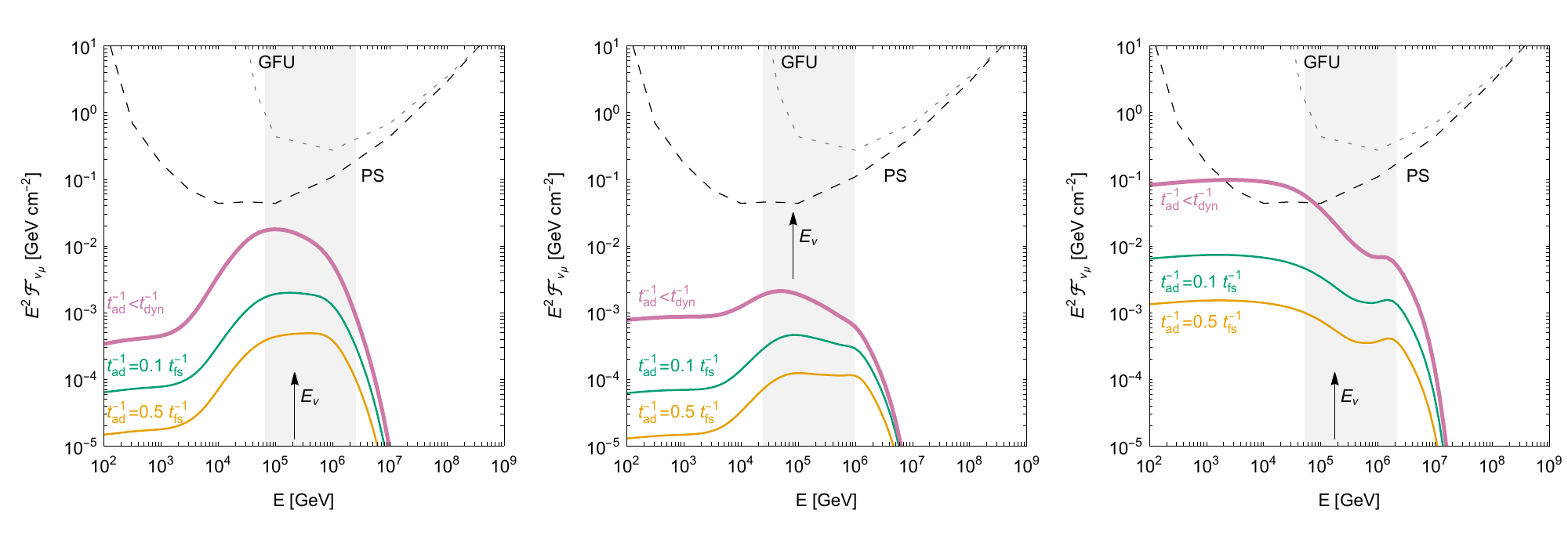} 
\end{center}
\caption{\label{fig:ad} Neutrino fluences for model M-X for \dsg , \fdr , and \aalc\ in the left, middle, and right panels respectively. In each panel, the different curves correspond to different adiabatic cooling rates, as given in the legends. Of these, the highest (reddish-purple) curve corresponds to \figu{mx} (see also that figure caption for details). }
\end{figure}

Since the adiabatic expansion rate depends on the nature of the radiation zone and its relationship to the accelerator, we do not include it in the main text, but  we discuss its effects on the neutrino fluence here. Adiabatic cooling can limit the calorimetric behavior of the system as the confined protons may lose energy by the expansion of the radiation zone faster than they can interact. Neglecting  adiabatic cooling is justified if the radiation zone is stable enough such that the adiabatic cooling rate is $t_{\mathrm{ad}}^{-1} \lesssim t^{-1}_{\mathrm{p\gamma}}$ (beyond the $p\gamma$ threshold). Fits of the BB radius, which typically decreases over time \citep{vanVelzen:2020cwu}, may support a stable production region. If, however, a non-relativistic outflow is indicative for the expansion of the system, then $t_{\mathrm{ad}}^{-1}$ is given by $t_{\mathrm{ad}}^{-1} \simeq 4/3 \cdot v/R \sim v/R \simeq 0.1 \, t^{-1}_{\mathrm{fs}}$ to $0.5 \, t^{-1}_{\mathrm{fs}}$ can be quite substantial for some models, see below.

First of all, note that adiabatic cooling affects the models differently. For model M-OUV, the radiation zone is optically thick, and adiabatic cooling will hardly have an effect on the (dominant) contribution from OUV interactions, see \figu{timescales_MOUV} ($t_{\mathrm{ad}}^{-1} < t^{-1}_{\mathrm{fs}} < t^{-1}_{\mathrm{p\gamma}}$ here). For model, M-IR, $t^{-1}_{\mathrm{ad}}$ is, because of the large radiation zone, potentially close to $t^{-1}_{\mathrm{dyn}}$ (for $t^{-1}_{\mathrm{ad}} \sim 0.1 \, t^{-1}_{\mathrm{fs}}$);  in either case, $t_{\mathrm{ad}}^{-1} \lesssim t^{-1}_{\mathrm{p\gamma}}$ here, see \figu{timescales_MIR}, so the effects of adiabatic cooling are expected to be small. We are therefore mostly targeting model M-X, see \figu{timescales_MX}, for which the adiabatic cooling rate could be potentially higher than the Bethe-Heitler cooling and photohadronic interaction rates, which can reduce the expected neutrino fluences.  

In \figu{ad}, we show the impact of adiabatic cooling on the neutrino fluences for model M-X for the three TDEs considered in this work and different adiabatic cooling rates corresponding to outflow models with $v=0.1 \, c$ and $0.5 \, c$ (leading to $t_{\mathrm{ad}}^{-1} \simeq  0.1 \, t^{-1}_{\mathrm{fs}}$  and $0.5 \, t^{-1}_{\mathrm{fs}}$, respectively). For comparison, the curves for $t_{\mathrm{ad}}^{-1} <  t_{\mathrm{dyn}}^{-1}$ (which is slower than the interaction rates, which means that adiabatic cooling is sub-dominant) are shown as well.  As it can be seen from the figure, the adiabatic cooling can suppress the neutrino fluence substantially because it may  dominate the cooling of the protons. We consequently note that especially for model M-X a stable radiation zone is required. Since $R \simeq R_{\mathrm{acc}}$ for that model, the acceleration zone should also not expand substantially (if it is identical to the radiation zone) or should release the protons at $R_{\mathrm{acc}}$ into a static radiation zone with $R \simeq const.$. We speculate that this could be \eg\ a sufficiently slow non-relativistic outflow with $v<0.1 \, c$, outflow-cloud interactions in a specific distance $R_{\mathrm{acc}}$ from the SMBH, an off-axis jet releasing the protons at $R_{\mathrm{acc}}$, \eg\ by advection, or a choked jet trapped in a quasi-static envelope.

\section{Consistency of the maximal proton energy}
\label{app:maxproton}

\begin{figure}[t]
\begin{center}
\includegraphics[width=0.3\textwidth]{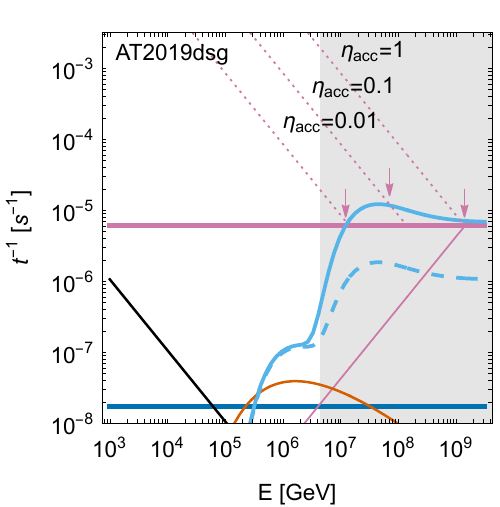} \hspace*{0.3cm}
\includegraphics[width=0.3\textwidth]{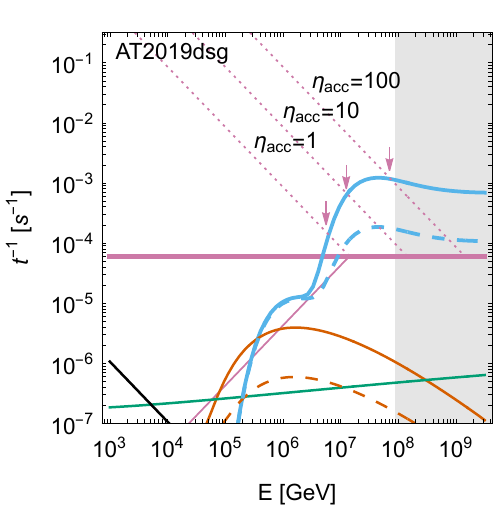} \hspace*{0.3cm} 
\includegraphics[width=0.3\textwidth]{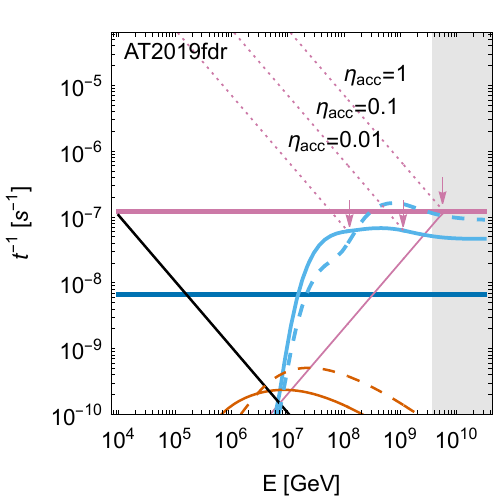} 
\end{center}
\caption{\label{fig:timescales_acc} Acceleration rates (dotted lines) for models M-X, M-OUV, and M-IR and for different values of the acceleration efficiency, $\eta_{\mathrm acc}$, are shown in the left, middle, and right panels, respectively, assuming that the acceleration and radiation zones are identical. The other curves are the same as the corresponding ones in \figu{timescales_MX}, \figu{timescales_MOUV}, and \figu{timescales_MIR}. The maximal proton energies for different $\eta_{\mathrm{acc}}$ at the BB peak time are marked by arrows, assuming that the fastest interaction or escape rate limits the maximal proton energy. Re-call that our model assumptions for $E_{p, \mathrm{max}}$ are given by the gray-shaded areas.}
\end{figure}

In the main text, we have parameterized the maximal proton energy $E_{p, \mathrm{max}}$. It is, however, an interesting question if $E_{p, \mathrm{max}}$ can be self-consistently described by a more realistic acceleration mechanism. A common parameterization for the acceleration rate, most appropriate for shock acceleration, is \citep{Hillas:1984ijl}
\begin{equation}
    t^{-1}_{\mathrm{acc}} = \eta_{\mathrm{acc}} \, \frac{c}{R_L} \propto \frac{1}{T_{\mathrm{cycle}}} \propto \frac{B}{E} \label{equ:acc}
\end{equation}
with an acceleration efficiency $\eta_{\mathrm{acc}}$ typically smaller than one depending on details of the model (e.g. shock velocity, compression rate); see \equ{larmor} for the definition of $R_L$. The maximal energy is then determined by the energy where radiative processes become faster than the acceleration.

We illustrate the acceleration rates and the processes determining $E_{p, \mathrm{max}}$ in \figu{timescales_acc}, assuming that a) the acceleration and radiation zones are identical, and that b) the interaction or escape rates limit the maximal energy (sometimes the cooling rates are used for this). With that we can now discuss the model-dependent implications for the acceleration efficiency for the shown examples by comparing the $E_{p, \mathrm{max}}$ obtained using \equ{acc} (arrows in \figu{timescales_acc}) with the assumptions for  $E_{p, \mathrm{max}}$ in the main text (gray-shaded areas in \figu{timescales_acc}):
\begin{description}
\item[M-X] The required acceleration efficiency is very low $\eta_{\mathrm{acc}} < 0.01$. The maximal energy will be likely determined by photohadronic interactions with the OUV target (at the BB peak). Therefore, it is plausible that acceleration and radiation zones are identical (e.g. outflow) or at a similar radius (e.g. off-axis jet), with only mild requirements for the acceleration.
\item[M-OUV] The required acceleration efficiency for the chosen value of $B$ is very large $\eta_{\mathrm{acc}}>100$ if the acceleration and radiation zones are identical. This means that the acceleration must happen in a zone with stronger magnetic fields which is also potentially more compact (unless a different mechanism, such as linear acceleration, is at work); examples could be corona or disk if high enough proton energies can be reached. 
\item[M-IR] The maximal energy matches $t^{-1}_{\mathrm{acc}}$ for $\eta_{\mathrm{acc}} \lesssim 1$, which means that the acceleration and radiation zones could be identical if the acceleration is efficient (e.g. hidden wind-like model, or outflow-cloud/circumburst material interactions); this does not exclude that the acceleration zone could be more compact with stronger magnetic fields (e.g. an off-axis jet). 
\end{description}

Note that there are other constraints on the acceleration region. For example, for shock acceleration triggered by the outflow, the shocks should not be radiation-mediated (which suppresses particle acceleration), see \eg\ \citet{Murase:2010cu}. The Thomson optical depth for outflow velocity $v$ can be estimated as
\begin{equation}
 \tau_T \simeq 0.07 \,  \left( \frac{M}{10^{7} \, M_\odot} \right) \left( \frac{R}{10^{15} \, \mathrm{cm}} \right)^{-1} \left( \frac{v}{0.5 \, c} \right)^{-1} \, ,
\label{equ:tt}
\end{equation}
which is typically much smaller than $c/v_s$ (with the shock velocity $v_s \sim v$ of the order of the outflow velocity). However, for more compact regions this constraint should be considered.


\end{document}